\documentclass[preprint,authoryear,12pt]{elsarticle}

\usepackage{amssymb}
\usepackage{rotating}
\usepackage{hyperref}

\def\cago{$^{12}$C$(\alpha,\gamma)^{16}$O }
\def\O16{$^{16}$O}
\def\C12{$^{12}$C}

\def\sun{\hbox{$\odot$}}
\def\la{\mathrel{\mathchoice {\vcenter{\offinterlineskip\halign{\hfil
$\displaystyle##$\hfil\cr<\cr\sim\cr}}}
{\vcenter{\offinterlineskip\halign{\hfil$\textstyle##$\hfil\cr
<\cr\sim\cr}}}
{\vcenter{\offinterlineskip\halign{\hfil$\scriptstyle##$\hfil\cr
<\cr\sim\cr}}}
{\vcenter{\offinterlineskip\halign{\hfil$\scriptscriptstyle##$\hfil\cr
<\cr\sim\cr}}}}}
\def\ga{\mathrel{\mathchoice {\vcenter{\offinterlineskip\halign{\hfil
$\displaystyle##$\hfil\cr>\cr\sim\cr}}}
{\vcenter{\offinterlineskip\halign{\hfil$\textstyle##$\hfil\cr
>\cr\sim\cr}}}
{\vcenter{\offinterlineskip\halign{\hfil$\scriptstyle##$\hfil\cr
>\cr\sim\cr}}}
{\vcenter{\offinterlineskip\halign{\hfil$\scriptscriptstyle##$\hfil\cr
>\cr\sim\cr}}}}}
 
\def\aj{AJ~}
\def\apj{ApJ~}
\def\aap{A\&A~}
\def\apjl{ApJ~}
\def\apjs{ApJS~}
\def\pasp{PASP~}
\def\aaps{A\&AS~}
\def\nat{Nature~}
\def\na{New Astron.~}
\def\araa{ARA\&A~}

\def\mnras{MNRAS~}
\def\aapr{A\&A~Rev.~}
\def\actaa{Acta Astron.~}
\def\zap{Z.~f\"ur~Astr.~}
\def\prl{Phys. Rev. Lett.~}
\def\prd{Phys. Rev. D~}
\def\jcap{J. Cosmol. \& Astropart. Phys.~}
\def\pasa{Pub. Astr. Soc. Australia~}
\def\memsai{Mem. Soc. Astron. Ital.~}

\journal{New Astronomy Reviews}

\begin{document}

\begin{frontmatter}


\title{The white dwarf luminosity function}

\author[1,2]{Enrique Garc\'\i a--Berro}
\ead{enrique.garcia-berro@upc.edu}
\author[3]{Terry D. Oswalt}
\ead{terry.oswalt@erau.edu}
\address[1]{Departament de F\'\i sica,
           Universitat Polit\`ecnica de Catalunya,
           c/Esteve Terrades 5,
           08860 Castelldefels, Spain}
\address[2]{Institut d'Estudis Espacials de Catalunya,
           c/Gran Capit\`a 2--4, Edif. Nexus 104, 
           08034 Barcelona, Spain}
\address[3]{Department of Physical Sciences,
            Embry-Riddle Aeronautical University,
            600 Clyde Morris Boulevard,
            Daytona Beach, FL 32114}
           
\begin{abstract}
White  dwarfs are  the final  remnants of  low- and  intermediate-mass
stars.  Their  evolution is essentially  a cooling process  that lasts
for  $\sim 10$~Gyr.   Their  observed  properties provide  information
about the history of the Galaxy, its dark matter content and a host of
other interesting  astrophysical problems.  Examples of  these include
an independent  determination of  the past history  of the  local star
formation  rate, identification  of  the objects  responsible for  the
reported microlensing events, constraints on the rate of change of the
gravitational  constant,  and  upper  limits to  the  mass  of  weakly
interacting massive particles.  To carry  on these tasks the essential
observational tools  are the  luminosity and  mass functions  of white
dwarfs, whereas  the theoretical tools are  the evolutionary sequences
of white dwarf progenitors, and  the corresponding white dwarf cooling
sequences. In particular, the observed white dwarf luminosity function
is the key  manifestation of the white dwarf  cooling theory, although
other  relevant   ingredients  are   needed  to  compare   theory  and
observations.   In this  review we  summarize the  recent attempts  to
empirically  determine the  white  dwarf luminosity  function for  the
different Galactic  populations. We also  discuss the biases  that may
affect its  interpretation. Finally,  we elaborate on  the theoretical
ingredients  needed  to model  the  white  dwarf luminosity  function,
paying  special  attention  to  the remaining  uncertainties,  and  we
comment  on  some applications  of  the  white dwarf  cooling  theory.
Astrophysical problems for which white  dwarf stars may provide useful
leverage in the near future are also discussed.
\end{abstract}


\end{frontmatter}

\section{Introduction}
\label{intro}

White dwarfs  are the  final evolutionary stage  of stars  with masses
less than $10\pm 2\, M_{\sun}$, though the upper mass limit is not yet
well known \citep{Rea99}.  \cite{vD04}, \cite{Sea04}, \cite{Mea05} and
\cite{Liea6} attempted to provide  observational limits on the maximum
progenitor mass.  Most white dwarfs are composed of carbon and oxygen,
but those with  masses less than $0.4\, M_{\sun}$ are  made of helium,
while  those more  massive than  $\sim  1.05\, M_{\sun}$  are made  of
oxygen  and   neon  \citep{GBI94}.   The  exact   composition  of  the
carbon-oxygen core critically depends on the processes occuring during
the  previous  Asymptotic  Giant   Branch  (AGB)  phase.   Theoretical
calculations show that the precise  ratio of carbon to oxygen depends,
to  a large  extent,  on the  competition between  the  \cago and  the
triple-$\alpha$ reactions  \citep{SA97}, on the particular  details of
the stellar  evolutionary codes  used to  compute the  pre-white dwarf
evolutionary   phases   \citep{R10},   on   the   adopted   convective
prescription  and  on the  choice  of  several other  physical  inputs
\citep{A05}.  In  a typical white  dwarf of $0.58\,  M_{\sun}$, oxygen
represents  62\% of  the total  mass, while  its concentration  in the
central  layers   of  the  white  dwarf   can  be  as  high   as  85\%
\citep{W92,SA97}.   In all  cases, the  core is  surrounded by  a thin
layer of pure helium with a  mass ranging from $10^{-2}$ to $10^{-4}\,
M_{\sun}$. Masses  larger than $10^{-2}\, M_{\sun}$  are not possible,
as this would lead  to helium ignition at the base  of the shell. This
region is,  in turn, surrounded by  an even thinner layer  of hydrogen
with mass  within the  range $10^{-4}$ to  $10^{-15}\,M_{\sun}$.  This
layer is  missing in $\sim 20\%$  of white dwarfs, and  determines the
basic chemical composition of their envelopes.

From  a phenomenological  point of  view, white  dwarfs with  hydrogen
spectral lines are  classified as DA.  The rest are  classified as DO,
DB,  DQ, DZ  and DC,  depending  on their  spectral features,  roughly
constituting a sequence of  decreasing temperatures \citep{Sea83}, and
have helium-dominated  atmospheres. It is well  established that white
dwarfs with  helium-rich atmospheres  are the result  of a  late shell
flash,  and   that  the  subsequent  evolution   results  in  distinct
atmospheric features.   For instance,  a dredge-up  episode is  at the
origin of  DQ white dwarfs,  while DZ white  dwarfs are the  result of
external  pollution by  metals. DO  and DB  white dwarfs  have surface
layers  which  are made  of  almost  pure  helium, only  differing  in
effective  temperature, which  determines  the corresponding  spectral
features, while  DC white dwarfs  are characterized by the  absence of
helium  spectral  lines  at  low  temperatures.   Although  our  basic
understanding of the physical mechanisms that lead to the formation of
the different white dwarf spectral  types is on solid grounds, several
details still need to be studied, as there is an interplay between the
mechanisms  operating  in  the  envelopes of  white  dwarfs,  such  as
gravitational  settling,  thermal   diffusion,  radiative  levitation,
convection at the H-He and He-core interfaces, proton burning, stellar
winds  and  mass  accretion   from  the  interstellar  medium.   These
interesting topics are,  however, out of the scope of  our paper --- a
good summary can be found  in \cite{CH2012}. Most of the observational
efforts to  empirically determine the white  dwarf luminosity function
have been done employing samples of  DA white dwarfs.  For this reason
we will focus primarily on work done on these types of white dwarfs.

White dwarfs have rather simple  mechanical structures. In fact, their
stability is mostly  provided by the pressure  of degenerate electrons
and  to remain  in hydrostatic  equilibrium  --- that  is, to  balance
gravity  with  the  pressure  gradient   ---  the  energy  release  of
thermonuclear  reactions  is  not  needed.   Because  of  this,  their
evolution can be described as a simple cooling process \citep{ME52} in
which the internal  degenerate core acts as a reservoir  of energy and
the outer non-degenerate  layers modulate the energy  outflow.  In the
simplest picture it  is assumed that the core is  isothermal, which is
justified by  the high  conductivity of  degenerate electrons  and the
thin  envelope.  Under  these  conditions energy  conservation may  be
written as:

\begin{equation}
L\approx-\frac{d U}{d t} = 
-\langle c_{\rm V}\rangle M_{\rm WD} \frac{d T_{\rm c}}{d t}
\end{equation}

\noindent where $U$ is the thermal content, $\langle c_{\rm V}\rangle$
is the  average specific heat, $T_{\rm  c}$ is the temperature  of the
nearly isothermal  core, and  $M_{\rm WD}$  is the  mass of  the white
dwarf.  Additionally, the  luminosity of the star  and the temperature
of  the isothermal  core  are,  to first  order,  related through  the
expression:

\begin{equation}
L  = f(T_{\rm c})M_{\rm WD}
\end{equation}

\noindent  where $f(T_{\rm  c})$ is  a function  which depends  on the
detailed thermal structure of the envelope.  This set of equations can
be  integrated,  provided that  $f(T_{\rm  c})$  is given.   A  simple
calculation indicates that  the cooling timescales of  these stars are
very long, $\sim 10$~Gyr, and  thus the white dwarf population largely
remains visible  throughout the cooling process  and retains important
information about the past history of the Galaxy.

In  particular, this  allows us  to derive  useful constraints  on the
stellar  formation rate  \citep{NS90,  DEA94, I95,  I01}  and the  age
\citep{WA87, GB88, W92, HA94, OSW96, BRL97, R00, H06} of the different
Galactic components: disk, halo as  well as open clusters and globular
clusters.  Moreover, it  has been conjectured that  those white dwarfs
that are  not detectable  could contribute  substantially to  the dark
matter content of our Galaxy. Specifically, surveys carried out by the
MACHO  team \cite{Aea95,  AEA97, AEA00}  suggested that  a substantial
fraction of  the halo dark  matter could be in  the form of  very cool
white dwarfs.  Since then, the  EROS \citep{LEA01, GEA02, TEA06}, OGLE
\citep{UEA94},  MOA \citep{MEA99}  and SuperMACHO  \citep{BEA05} teams
have  monitored millions  of stars  during several  years in  both the
Large Magellanic Cloud  (LMC) and the Small Magellanic  Cloud (SMC) to
search  for microlensing  events.  Most  of them  have challenged  the
results of  the MACHO experiment  --- see, for  instance, \cite{GCh04}
and references therein.   In addition, there have  been several claims
that white dwarfs could be the  stellar objects reported in the Hubble
Deep  Field  \citep{IA99,MM00,K05}.    However,  these  claims  remain
inconclusive for  lack of spectroscopic identifications  and confirmed
proper  motions.  The  Hubble Deep  Field--South has  provided another
opportunity to test  the contribution of white dwarfs  to the Galactic
dark  matter content.   In  particular, three  white dwarf  candidates
among  several faint  blue  objects which  exhibit significant  proper
motion have been found \citep{K05}.  They are assumed to belong to the
thick  disk  or  halo  populations.  If  these  are  spectroscopically
confirmed, it would imply that white  dwarfs account for $\la 10\%$ of
the  Galactic dark  matter,  which would  fit  comfortably within  the
results of the  EROS team.  All in  all, the study of  the white dwarf
population has  important ramifications  for our understanding  of the
structure and evolution of the Milky Way.

The fundamental  tool for studying  the properties of the  white dwarf
population is the white dwarf luminosity function, which is defined as
the number  of white  dwarfs per  cubic parsec as  a function  of unit
luminosity. The white  dwarf luminosity function not  only can provide
valuable information  about the  age, structure  and evolution  of our
Galaxy but it also provides an independent test of the theory of dense
plasmas  \citep{IEA97,  IE98a}.   Also,  it  directly  constrains  the
current death  rate of low-  and intermediate-mass stars in  the local
neighborhood  which,  in turn,  provides  an  important constraint  on
pre-white dwarf stellar evolutionary  sequences.  However, in order to
use the  white dwarf  luminosity function  to study  these interesting
astrophysical  problems, it  is necessary  to have  good observational
data, accurate stellar models, and reliable prescriptions to model the
population  components of  our Galaxy.   In this  paper we  review the
current knowledge  of the white  dwarf luminosity function,  from both
the observational and theoretical points of view.

The  paper is  organized as  follows.  Section~\ref{past}  reviews the
observational efforts, while in Sect.~\ref{theory} we provide an overview
of  the theoretical  models.  It  is followed  by Sect.~\ref{cooling},
where  we discuss  the current  state of  the art  of the  white dwarf
cooling  theory,  paying  special   attention  to  the  most  relevant
evolutionary phases.  For  the sake of brevity, we will  not review in
detail  the abundant  theoretical background,  but only  those salient
features  of the  theory needed  to model  the white  dwarf luminosity
function.   The  interested  reader  is  referred  to  \cite{R10}  and
\cite{ACIG} for detailed  discussions on topics such  as the so-called
convective coupling  or spectral evolution.  In  Sect.~\ref{inputs} we
discuss  other important  inputs necessary  to model  the white  dwarf
luminosity function.   In Sect.~\ref{applications}  we elaborate  on a
few  of  the  many  astrophysical  applications  of  the  white  dwarf
luminosity  function.  Section~\ref{future}  outlines the  foreseeable
future  research   in  the  field   from  both  the   theoretical  and
observational points  of view.  Finally,  Sect.~\ref{space} summarizes
in  a  comprehensive  way  the  most  relevant  observational  results
previously analyzed  in Sect.~\ref{past},  and lists  the ages  of the
Galactic  disk derived  from  the observed  luminosity functions.   We
conclude   with   a   general   summary,   which   is   presented   in
Sect.~\ref{conclusions}.

Before going into  details we would like to stress  that the selection
of papers  for explicit  citation may be  somewhat incomplete,  as the
field  is rapidly  evolving, and  moreover it  is the  product of  the
special research trajectory  of the authors.  We  apologize in advance
for any unintentionally missed references.


\section{The observed white dwarf luminosity function}
\label{past}

Over  fifty  years  ago  it  was first  recognized  that  the  coolest
(faintest) white dwarfs are remnants of  the earliest stars to form in
the Solar neighborhood \citep{Sch59}, and that cooling theory could be
used to  estimate the time  elapsed since star formation  commenced in
the Galactic disk \citep{ME52, VH68}.   Three decades later, the white
dwarf luminosity function helped resolve serious discrepancies between
the ages of the oldest stars in the Galaxy and the age of the Universe
implied  by the  Hubble  recession rate  of galaxies  \citep{W98,L99}.
Following  on  the heels  of  these  pioneering works,  several  other
investigations began to use white dwarfs as reliable cosmochronometers
to determine ages  of individual stars, binaries  and stellar clusters
--- see \cite{FBB01} for an excellent review of this topic.

The observed white dwarf luminosity function preserves a record of the
star formation  and death rate that  spans the history of  the Galaxy,
sets constraints  on the  quantity of its  local baryonic  matter, the
recycling  of material  to the  interstellar medium,  and encodes  the
kinematics of stellar  populations throughout the disk  and halo.  Its
uses  and inherent  limitations have  been  discussed in  a number  of
excellent papers \citep{W00, MR01, BLR01, HL03}, while several reviews
of  the theory  behind the  white dwarf  luminosity function  provided
essential  caveats and  context  for  its interpretation  \citep{DM89,
K02}.

Those new to the topic of the white dwarf luminosity function would do
well to start with the above  references as background.  Here we focus
on  a  few key  developments  leading  to  the  present state  of  the
empirical white dwarf luminosity function  and what can be expected in
the near future.  A comprehensive summary of all relevant work is well
beyond  the scope  of  this  review.  We  apologize  in  advance if  a
particular project of interest has been omitted.  However, each of the
works cited below contains an abundance of references and comparisons,
details on the  methods employed and, especially,  the myriad pitfalls
associated  with constructing  an  empirical  luminosity function  for
white dwarfs.

Any observational  study of the  white dwarf luminosity  function must
begin with a well-defined sample.  In the work summarized below, three
basic   approaches   have  been   used.    One   is  to   identify   a
magnitude-limited  sample  using  color index  selection  criteria  to
isolate the most  likely white dwarf candidates.   Another approach is
to use proper motions and color  indices to isolate nearby white dwarf
candidates by  their intrinsic low  luminosity, color and  high proper
motion.  Such  samples usually require  a weighting scheme  to correct
for the kinematic bias that causes fast-moving distant and slow-moving
nearby  objects  to be  undercounted.   The  best  approach is  to  do
straightforward  counts  in  a  large volume-limited  sample  that  is
demonstrably complete.  Unfortunately, this  is rarely possible due to
the low luminosity of white dwarfs,  confusion of the cooler ones with
lower main sequence stars and sub-dwarfs, the lack of spectra for many
objects, and the lack of precision trigonometric parallaxes.

In the works  described below all three of these  approaches have been
used.  The  discussion is  arbitrarily organized  into five  parts and
presented   in   rough   chronological   order   within   each   part.
Section~\ref{hot} outlines  several key attempts to  quantify the more
accessible hot  (bright) end of  the white dwarf  luminosity function,
which   constrains   the   current   white   dwarf   formation   rate.
Section~\ref{early} provides  an overview of early  (pre-2000) efforts
to  construct the  full  white dwarf  luminosity  function from  local
samples  of white  dwarfs.  Section~\ref{disk}  describes more  recent
progress towards a definitive luminosity  function for the local white
dwarf  sample.   This   work  has  tended  to   take  two  approaches:
construction  of complete  nearby  samples of  white  dwarfs that  are
effectively limited to the thin  disk and searches within large modern
surveys  that include  white dwarfs  from a  mix of  populations (thin
disk,   thick   disk,   halo).   Section~\ref{halo}   describes   very
preliminary attempts to construct the luminosity function specifically
for   the    halo   (spheroidal)    population   of    white   dwarfs.
Section~\ref{prospects} provides some concluding remarks on key issues
that still  need to be addressed,  as well as prospects  for improving
the observed white dwarf luminosity function in the near future.

\subsection{The hot end of the white dwarf luminosity function}
\label{hot}

The hot DA white dwarfs in the Palomar-Green (PG) Survey \citep{FLG86}
comprise a  magnitude-limited sample originally selected  primarily on
the basis of blue color criteria.   This sample was used to anchor the
hot  end  of one  of  the  first estimates  of  the  full white  dwarf
luminosity  function  \citep{WA87,  LDM88} discussed  below.   The  PG
sample of hot white dwarfs, i.e.,  those brighter than $M_v = 13$, was
analyzed  by  \cite{LBH05}  using  high  signal-to-noise  spectra  and
improved  atmosphere  model  fits  to   the  Balmer  lines  to  derive
temperatures,  gravities,  masses,  radii  and  cooling  ages.   Their
luminosity function, corrected for  incompleteness and weighted by the
$1/\mathcal{V}_{\rm max}$ method \citep{S68}  indicated that hot white
dwarfs comprise  about ten percent  of the white dwarf  space density.
This careful study is most  noteworthy for its examination of apparent
structure in  the white  dwarf mass  distribution and  for one  of the
first robust determinations of the formation rate of white dwarfs with
hydrogen-rich atmospheres.   The latter constrains the  formation rate
and  space density  of  planetary nebulae  and, consequently,  stellar
evolutionary models for progenitors less than about ten solar masses.

The  Kiso Schmidt  survey  of UV-excess  objects \citep{Kondo84}  also
proved to be a rich source of hot white dwarfs.  Using this magnitude-
and  color-limited sample,  supplemented  by  their own  spectroscopic
identifications, \cite{WD94}  published one  of the first  white dwarf
luminosity functions based on  this survey.  Their luminosity function
and space  density of hot white  dwarfs were found to  agree well with
that derived from the PG sample.

In a  preliminary analysis of  the Anglo-Australian Telescope  2dF QSO
Redshift Survey (2QZ) data, \cite{Vea05} identified $\sim 2,400$ white
dwarf candidates  at distances up  to 1~kpc above the  Galactic plane.
The main thrust of this work was  to measure both the scale height and
luminosity function  for hot white  dwarfs.  A white  dwarf luminosity
function  was presented  for  stars brighter  than  $M_v\sim 13$.   It
matches the early white dwarf  luminosity function of \cite{FLG86} and
has a  similar scale  height (200–-300~pc).   Notably, the  2QF sample
appears to be complete at the bright end, i.e., for $10 < M_v < 12.5$.

The hot end of the white dwarf luminosity function was evaluated using
the  Sloan Digital  Sky Survey  (SDSS) DR4  data by  \cite{K09}.  This
well-calibrated large sample included  almost 6,000 stars.  It enabled
a thorough examination of incompleteness and other systematic effects,
though they  did not attempt  to derive a  disk age or  space density.
They  also identified  a plateau  between $0.5  \leq M_{\rm  bol} \leq
3.8$.  \cite{hots} found that this plateau  could not be the result of
a sudden change in the white dwarf birth rate because it would also be
visible in the luminosity function  of helium-rich white dwarfs.  Once
stars with masses  smaller than the canonical limit  for the formation
of a  carbon-oxygen white  dwarf were  removed from  the observational
sample the agreement between theory and observation was nearly perfect
\citep{hots, 2015ASPC..493..343K}.

The  DA/non-DA ratio  as  a  function of  luminosity  is an  important
constraint  on the  evolutionary channels  that govern  the atmosphere
transformations and chemical composition  changes as the hottest white
dwarfs cool.  However, hot white dwarfs  are quite rare and this leads
to large errors in the observed DA/non-DA  ratio at the hot end of the
luminosity function.  Spectral  features tend to be weak  or absent in
cooler  white  dwarfs,  making  this  ratio  difficult  to  determine,
underscoring the importance of the hot white dwarf luminosity function
as a fundamental constraint on  the spectroscopic evolution of cooling
white dwarfs that influences the entire luminosity function.  However,
temperatures for some of the hot  DA white dwarfs can be overestimated
due to  unknown atmospheric metal  abundances; shifting them  to lower
luminosity bins  changes the shape  of the hot white  dwarf luminosity
function and the DA/non-DA ratio.

\cite{Limoges2010} presented  an analysis of  the hot DA and  DB white
dwarfs in  the Kiso  Schmidt survey,  using detailed  model atmosphere
fits to  the optical spectroscopic  data.  The resulting  $M_v$ values
were     compared      with     the     original      estimates     of
\cite{1994PhDT........26D},   which  were   obtained  from   empirical
photometric calibrations.  \cite{Limoges2010} found the two approaches
(spectra  and  photometry)  had  a  relatively  small  impact  on  the
calculated  luminosity  functions.    They  also  determined  separate
luminosity functions for  DA and DB stars and placed  a smaller number
of     stars     in     the    fainter     magnitude     bins     than
\cite{1994PhDT........26D}.  The luminosity functions, space densities
and  completeness determinations  they obtained  from the  Kiso sample
were found to be quite consistent with those published by \cite{LBH05}
for the  PG survey,  establishing the  hot end  of the  observed white
dwarf  luminosity   function  as  a  reliable   constraint  on  deeper
investigations.   Of  particular   note,  however,  \cite{Limoges2010}
discovered  several  unresolved   double-degenerate  binaries  in  the
sample,  raising  the possibility  that  other  undetected pairs  have
affected  estimates of  the  space  density of  white  dwarf stars  in
studies that do not use detailed atmospheric model fits to spectra.

\cite{K13} noted that removing low mass white dwarfs from their sample
increases the DA/non-DA  ratio at high effective  temperatures as well
as in the range known formerly as the DB gap.  He suggested that a new
SDSS white  dwarf catalog from  a later  data release could  provide a
large enough basis to begin to address these problems.

In summary, because  it fully incorporates new  theoretical models for
white dwarf  atmospheres, cooling, completeness, biases  and selection
effects,  the work  outlined above  --- see  additional references  in
\cite{2015ASPC..493..343K} --- can be  regarded as the most definitive
determination  of  the hot  white  dwarf  luminosity function  at  the
moment.  The PG  and Kiso surveys continue to  be valuable touchstones
for evaluating selection effects and  completeness of newer samples of
hot white  dwarfs.  Indeed, these  white dwarfs are often  embedded in
the larger newer studies.  Current samples are now so large that it is
no longer  observationally necessary  to consider the  hot end  of the
white dwarf luminosity function as  a separate project.  The following
sections  outline   attempts  to  fully  determine   the  white  dwarf
luminosity  function  to and  beyond  an  expected downturn  in  space
density at $M_{\rm bol} \sim 15$.

\subsection{Early work on the full white dwarf luminosity function for the
  Galactic disk}
\label{early}

To  the best  of  our knowledge,  the first  attempt  to construct  an
observational white  dwarf luminosity  function was  made in  the late
sixties  \citep{W67}.  Using  three datasets  \citep{L58, L63,  EG65},
\cite{W67} demonstrated that all closely followed the expected cooling
theory \citep{MR67}.  Assuming an age of 10~Gyr for the Galaxy, it was
estimated that white dwarfs as  faint as $M_{\rm bol}\sim 16.5$ should
have been found. At that time,  none fainter than $M_{\rm bol}\sim 15$
were  known.  In  hindsight,  it  was the  low  quantum efficiency  of
photographic  plates and  early electronic  detectors that  frustrated
early searches for faint (cool)  white dwarfs --- see \cite{Lea79} and
\cite{G86a, G86b}.

Many of  the early white  dwarf luminosity functions  were constructed
from Luyten's landmark proper motion surveys \citep{L63}.  Luyten used
second-epoch  red  plates taken  about  a  decade after  the  original
Palomar  Observatory Sky  Survey (POSS-I)  to measure  proper motions,
photographic magnitudes, and crude color classes for stars down to the
plate limit near  $m_{\rm pg}\simeq 21$ for roughly  two-thirds of the
sky.   With these  data  he identified  candidate  white dwarfs  using
so-called reduced  proper motion  diagrams, a technique  for isolating
stellar    populations    he    pioneered    \citep{L22}    ---    see
\cite{1972ApJ...177..245J} for an early  assessment of this technique.
Later, with  new high  quantum efficiency instrumentation  this sample
proved to contain many of  the previously ``missing'' cool faint white
dwarfs \citep{H86, OHL88}.

Over  thirty years  ago, it  was pointed  out by  \cite{L79} that  the
observed scarcity of white dwarfs of  very low luminosity could be due
to either large errors in the cooling  theory, or to the finite age of
the  Galaxy.  This  idea was  first  tested by  \cite{WA87}.  Using  a
sample   of  43   spectroscopically  identified   white  dwarfs   with
trigonometric parallax data  they concluded that the  absence of stars
in the lowest luminosity bin was statistically significant.  Expanding
on this  idea, the Luyten  Half-Second Catalog \citep{L75, L79}  --- a
subset of Luyten's POSS-I proper motion survey —-- was used for a more
detailed  analysis   of  the   white  dwarf  luminosity   function  by
\cite{LDM88}.   A sample  of  mostly  cool white  dwarfs  with $\mu  >
0.8$~sec~yr$^{-1}$ and $M_v  > 13$ was selected  to better distinguish
nearby  low  luminosity white  dwarfs  from  high velocity  background
stars.  To maximize completeness, hot white dwarfs were added from the
color-selected PG  sample of \cite{FLG86}.  There  was initial concern
that  the \cite{LDM88}  sample  was not  complete,  especially at  the
critical faint end that constrains  the age of the Galaxy \citep{IL89,
OSW96, Fea01}. In a more complete sample, the position of the downturn
would  move  to   fainter  magnitudes  and  hence   the  Galactic  age
determination  would  increase.   However,  many of  the  fields  were
re-examined  by  \cite{Mo00}  using  new POSS-II  plates  and  it  was
concluded that the LHS sample is  roughly 90 percent complete over the
magnitude  and proper  motion limits  used in  \cite{LDM88}, when  the
scale  height  of the  Galaxy  is  taken  into account.   This  sample
provided the  first reliable  estimate for the  local mass  density of
white dwarfs and a minimum age for the Galactic disk of $\sim 9$~Gyr.

The Luyten  POSS-I survey was  also used  by \cite{OSW96}, but  with a
different  approach.   A  large  sample of  wide  binaries  containing
spectroscopically identified white dwarfs with much fainter magnitudes
and much  smaller proper  motions were  selected.  This  permitted the
sample to be as  deep as possible and also to  include any nearby slow
moving white dwarfs  that would have been  overlooked by \cite{LDM88}.
Corrections for incompleteness in the sample were made by constructing
star  counts vs.  both magnitude  bin and  proper motion  bin.  $BVRI$
photometry  was  used to  estimate  the  white dwarf  luminosities  by
interpolating  within  the  grids  of  hydrogen-rich  and  helium-rich
atmosphere  cooling  models  by   \cite{W95}.   The  model  chosen  to
determine the luminosity for each of the $\sim 50$ stars in the sample
was determined  by the dominant  constituent seen in its  spectrum.  A
mass of  $0.6\, M_{\sun}$  was assumed  for stars  without independent
mass determinations.  The uncertainty in composition and the empirical
dispersion  in   the  white  dwarf  mass   distribution  published  by
\cite{BSL92} were included in the error analysis.  This study revealed
that the downturn was less  steep than found by \cite{LDM88}, implying
a somewhat older minimum age of $\sim 9.5$~Gyr for the Galactic disk.

\cite{LRB98}  obtained optical  and infrared  data for  the sample  of
\cite{LDM88}.   Using  model  atmospheres   by  \cite{BSW95}  and  new
trigonometric parallaxes,  they recomputed the white  dwarf luminosity
function.  The result yielded only slight differences in the shape and
a modest  increase in  disk age  from \cite{LDM88}.   Incorporation of
improved  trigonometric   parallaxes  (the  dominant   uncertainty  in
luminosity   estimates)  and   additional   leverage  on   atmospheric
composition provided by the infrared data for the cooler objects, made
the \cite{LRB98}  a more definitive  determination of the  white dwarf
luminosity function than prior work.

\cite{KHH99}  constructed  one of  the  first  white dwarf  luminosity
functions not based on the Luyten proper motion survey.  Digital scans
of $\sim 300$ SuperCOSMOS ESO/SERC  plates were used to identify white
dwarf candidates via reduced proper  motion diagrams.  Both the proper
motion and photographic  color estimates were much  improved over what
Luyten could  do with  the POSS plates.   More importantly,  a special
effort was made to assess the incompleteness of the sample of 58 white
dwarf candidates that were identified.  It  is one of very few samples
shown  to  pass the  $\langle  1/\mathcal{V}_{\rm  max}\rangle =  0.5$
completeness   test   \citep{Sch75}.    The   atmosphere   models   of
\cite{BSW95}  were  used  to  estimate  the  white  dwarf  candidates'
luminosities.  The magnitude  and proper motion limits  of this survey
significantly exceeded that of \cite{LRB98}, enabling the detection of
intrinsically fainter  stars.  The  downturn at the  faint end  of the
white dwarf luminosity function was found to be more gradual than most
earlier surveys,  implying a  minimum disk age  of $\sim  9.5$~Gyr, in
accord with the findings of \cite{OSW96}.

The   white   dwarf   luminosity    function   was   re-evaluated   by
\cite{1997PhDT........16S}  and   summarized  by   \cite{Sea03}  using
much-improved  photometric  and spectroscopic  data  for  a sample  of
Luyten white dwarfs in wide binaries more than three times larger than
used  in  \cite{OSW96}.   Using  the  same  incompleteness  correction
strategy and  the $1/\mathcal{V}_{\rm  max}$ methodology,  the revised
luminosity  function,  space  density  and disk  age  were  unchanged.
However, the larger  sample improved the precision of each  by about a
factor of two.

The  above   studies  marked  important  stepping   stones  towards  a
definitive white  dwarf luminosity  function.  Because  of its  tie to
good  trigonometric  parallaxes  and careful  fitting  of  atmospheric
models  to   the  white  dwarf  spectral   energy  distributions,  the
luminostiy  function  of  \cite{LRB98}  probably  is  the  best  early
benchmark.   However,  the  work   by  \cite{OSW96}  and  \cite{KHH99}
convincingly demonstrated that deeper surveys were needed to delineate
the faint end  of the white dwarf luminosity  function that constrains
the age  of the  Galaxy and  to identifying the  fraction of  each bin
contributed by  its halo  components.  These  issues spurred  the more
recent work  using large  modern surveys, which  are addressed  in the
next two sections.

\subsection{The disk white dwarf luminosity function: balancing quality
  and quantity}
\label{disk}

Newer work on the white dwarf  luminosity function has tended to focus
on obtaining ``quality'', i.e.  well-vetted complete samples of nearby
white  dwarfs  drawn  from  a variety  of  sources,  or  ``quantity'',
i.e.  large samples  of white  dwarfs gleaned  from huge  surveys with
well-quantified completeness  characteristics.  We begin  this section
with some examples of the first type.

Only  within the  last decade  or so  has the  census of  nearby white
dwarfs  grown  large  enough  to  seriously pursue  the  best  way  to
determine  a  luminosity  function:  direct  star  counts  by  volume.
\cite{HSO08} used  the Catalog  of Spectroscopically  Identified White
Dwarfs  \citep{McS99} to  identify a  sample  of well  over 100  white
dwarfs  with  high quality  spectra,  photometry,  proper motions  and
parallaxes likely to  be within 20~pc of the Sun.   Their subset of 44
white  dwarfs within  13~pc of  the Sun  was shown  to be  essentially
complete and  the 20~pc sample  as a whole was  shown to be  almost 80
percent complete.   Thus, simple star  counts could be used  to derive
the  space   density.   This  avoids   the  necessity  of   using  the
$1/\mathcal{V}_{\rm  max}$ method  or  more sophisticated  statistical
methods, all  of which are  susceptible to small  number fluctuations,
observational biases, and/or unproven completeness --- see \cite{WO98}
for a discussion of these problems.

The  kinematical   properties,  spectroscopic  subtypes   and  stellar
population subcomponents  of the 20~pc  local white dwarf  sample were
evaluated  by \cite{S09}.  Virtually the  entire sample  was found  to
belong to  the thin disk component  of the Galaxy.  This  local sample
contained   not   a   single   interloping   member   of   the   halo.
\cite{2014AJ....147..129S}  confirmed  these  results in  an  expanded
sample of over  200 white dwarfs within 25~pc.   The completeness’s of
these samples  were shown  to be  well-behaved functions  of distance,
from nearly 100\% at 13~pc, to 85\%  at 20~pc, to 60\% at 25~pc.  They
provide useful benchmarks against which  to assess and compare samples
comprised of more distant white dwarfs.

Other groups have  been working hard to extend and  complete the local
white dwarf  sample.  \cite{2012ApJS..199...29G} performed  a detailed
photometric  and/or  spectroscopic  analysis   of  every  white  dwarf
suspected to lie within 20~pc of the Sun.  The sample was mostly drawn
from the 20~pc list given  in \cite{S09}.  The sample completeness was
estimated to be about 90\%.  While  it agrees well with the work cited
above in the low luminosity bins, at higher luminosities (white dwarfs
hotter than about  $T_{\rm eff} = 12,000$~K)  an apparent over-density
of a factor of two relative to these other luminosity functions was of
concern.  Most likely,  this was due to the small  number of stars ($<
10$) in the  brighter bins, a hypothesis that can  only be tested with
much larger samples.

\cite{HOS16} completed a  new analysis of the  25~pc sample, expanding
the  count of  spectroscopically identified  white dwarfs  by about  a
factor of  two relative to  the original 20~pc sample.   This expanded
sample provided  evidence that  single white dwarfs  are significantly
more represented  than those with  one or more  companions, suggesting
that some  companions are lost  to mergers or escape  during post-main
sequence  evolution.   In  addition,  this study  provided  the  first
estimate of the white dwarf birthrate as a function of time from $\sim
8$~Gyr ago to  the present, indicating that the  present production of
white dwarfs is a factor of two  to three higher than the average over
this period.   If mergers contributed  a significant component  to the
white dwarf population  or the birthrate has  changed significantly in
the Galaxy  these effects will  need to be considered  in interpreting
the white dwarf luminosity  function.  These concerns notwithstanding,
the \cite{HOS16}  white dwarf  luminosity function obtained  by simple
number  counts is  in good  agreement with  prior work  and should  be
considered a more fundamental determination.

In parallel  with studies  focused on the  luminosity function  of the
nearby  sample  of  white  dwarfs,  much  effort  has  been  put  into
quantifying it using huge samples of white dwarf candidates drawn from
large new  surveys.  We now  summarize a  few of the  most significant
``quantity-based projects''.

The  Luyten sample  still has  potential  to improve  the white  dwarf
luminosity  function.   \cite{SG02,  SG03} used  the  NLTT  positions,
magnitudes  and colors  to  cross-correlate to  the  2MASS and  USNO-A
survey data  to construct a catalog  of white dwarf candidates  in the
overlap  of regions  covered by  these surveys.   The improvements  in
proper motion estimates and photometric indices derived from the 2MASS
$J$-band  and  estimated photographic  $V$  magnitudes  from the  NLTT
permitted construction of reduced  proper-motion diagrams that cleanly
separate the main sequence, subdwarfs,  and white dwarfs.  The task of
obtaining    high    quality   spectroscopic    identifications    and
cross-correlated $ugriz$  photometry for  nearby cool white  dwarfs in
the NLTT catalog  was undertaken by \cite{KV06} but, to  date, it does
not appear that a white dwarf luminosity function has been constructed
for this spectroscopically identified portion of the sample.

\begin{figure}[t]
\centering
\includegraphics[width=0.8\textwidth,clip]{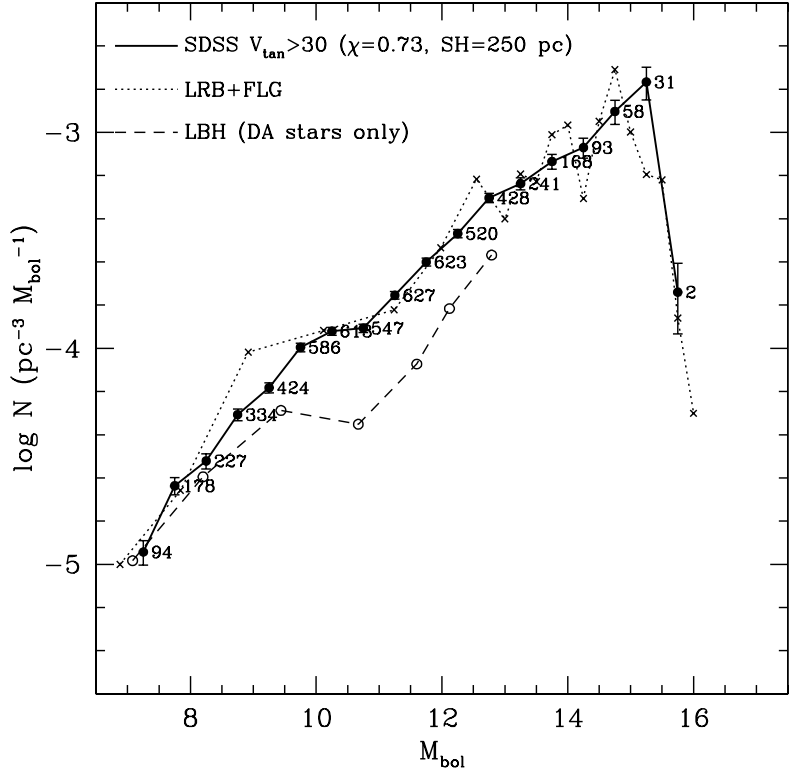}
\caption{Luminosity  function of  disk white  dwarfs derived  from the
  Sloan Digital Sky Survey (SDSS),  from \cite{Hea06}, compared to the
  luminosity functions of \cite{LRB98} and \cite{LBH05}.}
\label{fig:disk}
\end{figure}

\cite{Hea06}  constructed  a  white dwarf  luminosity  function  using
proper motions based  on comparison of positions between  the SDSS and
USNO   surveys,  high   quality  $ugriz$   photometry,  and   improved
atmospheric models.   As shown  in Fig.~1,  the resulting  white dwarf
luminosity function is  surprisingly smooth and drops  off abruptly at
$M_{\rm bol} = 15.3$.  Because of the scarcity of stars in bins beyond
the downturn, no attempt was made to derive the age of the disk or the
space  density of  white dwarfs.   Both \cite{Hea06}  and \cite{Kea06}
attempted to address the incompleteness of the SDSS white dwarf sample
as well  as the effects of  contamination by other types  of stars and
unresolved components, but in fact the SDSS does not adequately sample
the nearby  population of  white dwarfs due  to the  original survey’s
bright limit,  nor does it go  deep enough to delineate  the downturn.
Incidentally, the  \cite{Vea05} hot  white dwarf  candidates mentioned
above have been shown to be from the same parent population sampled in
the  northern  hemisphere  by  \cite{Hea06}.   Thus,  it  provides  an
important supplement to  the incompleteness of the SDSS  at the bright
end of the luminosity function.

\cite{Limoges2013} examined the white dwarf population within 40~pc of
the Sun using a spectroscopic survey of northern hemisphere candidates
from  the  SUPERBLINK proper  motion  database.   The expanded  survey
subsequently evaluated  by \cite{Limoges2015}  is between 66  and 78\%
complete.  It contains almost 500  white dwarfs, an order of magnitude
more than the original \cite{LDM88}  luminosity function.  Using a set
of  homogeneous model  atmospheres, they  found an  unexpectedly large
fraction of massive white dwarfs.   These less luminous objects at the
faint end  of the  luminosity function were  often missed  by previous
surveys.   Like   \cite{2012ApJS..199...29G}  the  disk   white  dwarf
luminosity function  obtained from  this sample  also has  an apparent
excess of hot white dwarfs, most likely due to contamination by non-DA
white  dwarfs as  in  the \cite{K13}  sample.  The  \cite{Limoges2015}
luminosity function  is the only  one based on a  40~pc volume-limited
sample.  However, trigonometric parallaxes are still needed to clearly
define the shape of the faint end of the luminosity function where the
disk and halo white dwarfs mingle.

\cite{40pc} simulated  the \cite{Limoges2015} survey.  They  were able
to reproduce the  observed white dwarf luminosity  function quite well
and  showed the  sample  completeness is  typically  $\sim 80$\%,  for
$M_{\rm bol}  < 16$, beyond which  it drops rapidly below  20\%.  They
also  demonstrated  that  the  downturn  in  the  observed  luminosity
function  located at  $M_{\rm  bol}\sim 15$  is statistically  robust.
Using  new progenitor  evolutionary  models and  cooling sequences  to
model  the  white  dwarf  luminosity  function  constructed  from  the
\cite{Limoges2015}  sample,  they  determined  the age  of  the  Solar
neighborhood to be about $8.9\pm 0.2$~Gyr,  about a factor of two more
precise than  the best  prior age  determinations.  This  estimate was
shown not  to depend significantly  on the  slope of the  initial mass
function or the adopted  initial-to-final mass relationship.  However,
the  peak in  the luminosity  function  was found  to be  shaped by  a
steeper initial-to-final  mass relation  for progenitor  masses larger
than  about  $4\,  M_{\sun}$.   An apparent  bump  in  the  luminosity
function near  $M_{\rm bol}\sim  10$ was found  to be  significant and
most likely the result of a  recent burst of star formation about $0.6
\pm 0.2$~Gyr ago  that continues to the present era.   This is roughly
in accord with the empirical  results of \cite{HOS16} derived from the
25~pc sample.

Another  significant step  towards a  definitive observed  white dwarf
luminosity function is the study of  DA white dwarfs identified in the
LAMOST  (Large Sky  Area Multi-Object  Fiber Spectroscopic  Telescope)
Spectroscopic  Survey   of  the   Galactic  anticenter   (LSS-GAC)  by
\cite{2015MNRAS.450..743R}.  Their  study followed a  well-defined set
of criteria  for selecting  targets for  observations, in  contrast to
other large  surveys with  target selection algorithms  complicated by
other  scientific  goals  that  make  it  difficult  to  quantify  the
observational  biases influencing  the observed  populations of  white
dwarfs.   Even  so,  \cite{2015MNRAS.452..765G}  determined  that  the
\cite{2015MNRAS.450..743R} survey's incompleteness is too large in the
faintest bins to confidently extend the luminosity function beyond the
downturn.

To  summarize this  section,  the luminosity  function  of disk  white
dwarfs has been the subject of a number of studies employing different
approaches.  Over  the luminosity  bins in common,  all of  them agree
within  the uncertainties  quoted by  each of  the surveys.   All have
consistently  found that,  beginning with  its bright  (hot) end,  the
white   dwarf  luminosity   function   increases  monotonically   with
increasing  bolometric magnitudes  at a  nearly constant  slope (which
stems from  the cooling law),  and terminates with an  abrupt downturn
near $M_{\rm  bol}\sim 15$  (a consequence  of the  finite age  of the
Galactic  disk).   Enticing but  very  preliminary  evidence for  fine
structure in  the white  dwarf luminosity  function suggests  that the
star formation rate  in the solar neighborhood has  not been constant.
At present, because of its large sample size and all the work that has
been put  into quantifying  its systematic errors,  incompleteness and
biases, the SDSS-based \cite{Hea06} white dwarf luminosity function is
the best benchmark against which newer studies should be compared.

\subsection{In search of the halo white dwarf luminosity function}
\label{halo}

It  was early  noted  \citep{Sch59} that  the  white dwarf  luminosity
function could give useful constraints on  the age and mass density of
the  Galactic halo.   The importance  of detecting  and characterizing
this population of the Galaxy's oldest  stars to such questions as the
age of the  halo and to the  nature of detected MACHO  events has been
reviewed by \cite{FBB05}.  In one of the first attempts, \cite{LDM88b}
used just six stars with high space motions to construct a preliminary
halo  white  dwarf  luminosity  function.   Since  then,  only  modest
increases have  been achieved  in the number  of confirmed  halo white
dwarfs,  because they  comprise such  a  rare component  of the  solar
neighborhood.   In addition,  the detection  of halo  white dwarfs  is
hampered  by  the  difficulty  of obtaining  their  three  dimensional
kinematical properties.   Sophisticated neural network  techniques may
prove  to  be  useful  in  addressing  these  problems  \citep{Tea98}.
However, radial velocities  are difficult to measure  for single white
dwarfs because  of their sizable gravitational  redshifts and frequent
lack  of measurable  absorption lines,  making full  three-dimensional
kinematics  hard to  come by.   This, in  turn, results  in difficulty
distinguishing between  thick disk and halo  white dwarfs \citep{PN03,
PN06}.

The  theoretical   prediction  \citep{H99,  SJ99}   and  observational
confirmation \citep{Ha99,F04} of depressions in the near-IR spectra of
very cool white  dwarfs underscored the need for  improvements in cool
atmosphere models as  well as better observational  data.  Below about
4,500~K,  the  infrared  colors  of white  dwarfs  with  hydrogen-rich
atmospheres become bluer  as cooling progresses, due  to broad opacity
sources  such   as  collisionally   induced  absorption   by  hydrogen
molecules.  This immediately  caused concern that the  fainter bins in
some  of  the white  dwarf  luminosity  function determinations  might
contain  stars   whose  luminosities  had  been   based  on  erroneous
photometric parallaxes or atmospheric models.

The discovery of ``cool blue degenerates'' sparked renewed interest in
finding old halo white dwarfs in the solar neighborhood.  For example,
the  large space  densities reported  by \cite{IA99}  and \cite{Oea01}
ignited  a flurry  of excitement  at the  prospect that  ancient white
dwarfs might comprise most or all of the Galaxy’s dark matter content.
Most of  these objects  were subsequently  shown to  be the  result of
misidentifications  of  thick  disk   white  dwarfs  and/or  stars  of
indeterminate proper  motion --- see \cite{Rea01},  \cite{Siea02}, and
references therein.  Ultimately, no real improvement in the halo white
dwarf luminosity function was obtained.

In  an attempt  to find  halo  white dwarfs,  \cite{MS02} applied  the
reduced proper  motion technique to  a ``pencil beam'' sample  of over
800 faint stars ($B < 22.5$)  with proper motions of high precision in
a 0.3 square  degree field near the north Galactic  pole.  Eight white
dwarf candidates were  identified in this field.   Taking into account
the narrowness  of the  field and  the much  smaller areal  density of
nearby white dwarfs across the  sky, they concluded that a substantial
faint  population of  white dwarfs  may  extend well  above the  scale
height of the  Galaxy without significantly affecting  the local white
dwarf space density.  Their maximum  likelihood method was shown to be
far  less  sensitive to  small  number  fluctuations that  affect  the
$1/\mathcal{V}_{\rm   max}$   method   used  by   nearly   all   other
investigations of the white dwarf luminosity function.

A robust white  dwarf luminosity function for the  halo should contain
at least as many stars ($\sim 50$) as the early white dwarf luminosity
functions for the disk.  This immediately brings to mind large surveys
such as  the SDSS.  However,  despite a  much larger sample  size, the
original SDSS  does not probe  as deeply as the  original \cite{LDM88}
study.  Among  spectroscopically identified white dwarfs  in the SDSS,
\cite{HWW07} showed that the sample is complete only between $16 < g <
18$.  This is not surprising, since stars were not the primary targets
of  the  SDSS.  This  sample  is  therefore  subject to  very  complex
selection effects that are strong functions of magnitude, area, color,
position    in    the    sky     and    other    factors    ---    see
\cite{2015MNRAS.452..765G}.   Despite  these  obstacles,  \cite{HWW07}
were able  to derive improved estimates  for the DA white  dwarf space
density  and formation  rate that  are  in good  agreement with  prior
values.

The SuperCOSMOS and RECONS surveys  \citep{Hea04, Sea05} were aimed at
detecting halo white dwarfs.  They identified $\sim 104$ new candidate
white  dwarf stars,  almost two  orders of  magnitude larger  than the
first samples  used to determine  the white dwarf  luminosity function
several decades  ago.  \cite{RH11}  used this  sample to  estimate the
space density of white dwarfs in the halo.  Although the majority lack
spectroscopic confirmation, the colorimetric and reduced proper motion
criteria used to identify candidates  were shown to be reliable; known
white dwarfs in the sample were readily identified.  Importantly, this
group introduced a new technique  to convincingly distinguish the disk
and halo white dwarfs for the  first time, extending the search to 1.0
and  2.5  magnitudes deeper,  respectively,  than  the SDSS  study  of
\cite{Hea06}.  Their results  confirm the location of  the downturn in
the disk white  dwarf luminosity function near $M_{\rm  bol} = 15.75$.
They also  concluded that the different  kinematic populations overlap
so  seriously beyond  the  peak  in the  disk  white dwarf  luminosity
function that traditional approaches to constructing one cannot render
a more accurate  (thin) disk age at  the present time and  only a very
preliminary halo age of 11--12~Gyr.

In  view of  the  large sample  size  and care  taken  to account  for
incompleteness,  quantify  selection  effects and  unravel  population
mixing, the \cite{RH11} white dwarf  luminosity functions for the thin
disk, thick disk  and halo are the best currently  available for these
population components.

\subsection{Prospects for improving the disk and halo white dwarf
  luminosity functions}
\label{prospects}

The various  studies outlined  above have  determined the  white dwarf
luminosity function  for thin  disk members  to a  precision of  a few
percent for  stars brighter  than the  downturn near  $M_{\rm bol}\sim
15$.  Most  of the  ongoing and  planned surveys  seek to  find enough
white dwarfs to populate the  lowest luminosity bins of the luminosity
function where the disk and  halo populations are currently hopelessly
mixed, and mired  in small number statistical  uncertainties.  Many of
these white  dwarfs are likely  be the ``cool blue  degenerate stars''
noted above, if they have hydrogen-rich atmospheres.  Whether they are
the dominant component  of the halo white dwarf  population remains to
be  seen.  This  has  been  a topic  of  considerable  debate ---  see
\cite{Tea07}, \cite{Tea08} and \cite{Tea10}.

\cite{2015MNRAS.452..765G} pointed the  way to better use  of the SDSS
for constructing the white dwarf  luminosity function.  Using the SDSS
DR10, they developed a selection method for white dwarfs that reliably
identified white  dwarf candidates  based on  SDSS colors  and reduced
proper  motion.  From  a large  sample of  spectroscopically confirmed
white  dwarfs and  known contaminants  (i.e., non-white  dwarfs) drawn
from the SDSS  DR7 they devised a method of  computing the probability
of being a white dwarf for any object having only multiband photometry
and  proper motion  data.   The spectroscopic  sample  was limited  to
bright objects ($g  < 19$) for which reliable proper  motions could be
obtained from prior photographic plates. Applying the technique to the
SDSS  DR10   photometric  catalogue,   they  selected   $\sim  23,000$
high-confidence white dwarf candidates,  of which $\sim 14,000$ lacked
spectra.  On  average, the sample  was found to  be only about  40 per
cent complete for white dwarfs hotter than $T_{\rm eff}\simeq 7,000$~K
and  brighter than  $g  \simeq 19$.   While they  did  not attempt  to
construct a white dwarf luminosity function, their results underscored
both the remaining potential of the  SDSS for improving it, as well as
the continuing need for follow-up spectroscopy.

Almost 40,000 white dwarfs  now have been spectroscopically identified
in  the various  extensions  of the  SDSS, up  through  its 12th  Data
Release \citep{2015ApJS..219...12A}. Among  these are several thousand
new white  dwarfs identified by  \cite{DR10} in the SDSS  DR10.  Using
the best SDSS spectra and  the latest atmospheric models they computed
temperatures,  gravities,  and   atmospheric  abundances  for  several
thousand hydrogen atmosphere white  dwarf stars (DAs), several hundred
helium atmosphere white dwarf stars (DBs),  as well as dozens of white
dwarfs  with  metallic  lines  (DZs)  and  white  dwarfs  with  carbon
dominated  spectra (DQs).   They also  constructed the  best currently
available  white dwarf  mass  distribution using  model  fits to  high
quality  SDSS spectra  for  $\sim  6,000$ DAs  and  corrected for  the
observed volume via the $1/\mathcal{V}_{\rm max}$ method.

The  \cite{DR10} identifications  reach  to $T_{\rm  eff} =  5,000$~K,
although in this regime the sample  is certainly not complete, as they
relied on proper motion measurements  (known to be incomplete below $g
\sim  21$)  to  distinguish  between  cool DCs  and  BL  Lac  objects.
However,  this  huge  increase  in  the  number  of  spectroscopically
confirmed white  dwarfs is important  because it enabled  discovery of
many rare objects such as massive white dwarfs, magnetic white dwarfs,
and He-dominated objects with oxygen lines, unresolved binaries,\ldots
Of special note, they compiled a  list of nearly 100 white dwarf stars
with masses above  $1\, M_{\sun}$ and found that  the volume corrected
distribution is  inhomogeneous.  If confirmed, this  may imply mergers
are a significant contributor to the white dwarf luminosity function.

While  neither  presented  new   determinations  of  the  white  dwarf
luminosity  function, the  \cite{2015MNRAS.452..765G} and  \cite{DR10}
studies are important steps to this goal and their strategies could be
applied to even larger future samples.  Compared to previous work they
followed much more well-defined criteria for selecting targets.  Their
assessments also  revealed that  the incompleteness at  the bolometric
magnitudes typical of the downturn in the luminosity function is still
large, and thus  more work with still larger samples  will be required
to  derive  a  reliable   luminosity  function  at  faint  magnitudes.
\cite{Sayres2012}  demonstrated  that   a  multi-survey  approach  can
improve detection of  nearby faint white dwarfs of  low proper motions
and rejecting contaminating populations of stars.

In  summary,  substantial improvement  in  the  present state  of  the
observed  white   dwarf  luminosity  function  will   require  several
advancements in  both the  quality and  quantity of  the observational
data, as  well as  improvements in  the models  used to  construct and
interpret it:

\begin{enumerate}
\item An  ultra large sample of  white dwarfs, on the  order of $10^5$
  stars.
\item Precise parallaxes, substantially better than 1 mas.
\item Precise proper motions, substantially better than 1 mas/yr.
\item High  quality photometry  for a  magnitude-limited sample  to at
  least $g\sim 21$.
\item  Spectroscopic  identifications  of  sufficient  resolution  for
  velocity determinations.
\item Improved atmospheric models for very cool white dwarfs.
\item Improved spectral evolutionary models.
\item Better categorization and treatment of selection effects.
\item Quantification  of the effects  of unresolved binaries  and high
  mass white dwarfs.
\end{enumerate}

Some of the above requirements can  be met by existing surveys such as
the expanded SDSS.   Until Gaia, however, it is unlikely  that a truly
definitive  white  dwarf  luminosity  function  for  either  the  disk
population  components  or  halo  will be  achieved,  primarily  since
precision parallaxes  and proper motions  are essential to  proper bin
assignment and resolution  of the various populations that  mix at the
faint end of the observed white dwarf luminosity function.  This is an
intractable problem for  the current surveys.  In  short, the downturn
in the disk luminosity function for  now is ill determined below about
$10^{-4}\, L_{\sun}$.  Further, because  they comprise a tiny fraction
of the local  population, a complete (or at least  a very well-behaved
incomplete)  sample  of  white  dwarfs  needs  to  be  constructed  to
distances  approaching  1 kpc  in  order  to capture  a  statistically
significant number of faint  halo white dwarfs.  \cite{TGB05} estimate
that  Gaia will  find 250,000  to 500,000  white dwarfs  --- see  also
\cite{Carrasco}.  This  will open a whole  new era of research  on the
white dwarf luminosity function.


\section{Theoretical models of the white dwarf luminosity function}
\label{theory}

The white dwarf luminosity function can be formulated as

\begin{eqnarray}
n(L)\propto\int^{M_{\rm s}}_{M_{\rm i}}\,\Phi(M)\,\Psi(T-t_{\rm cool}
(L,M)\nonumber\\
-t_{\rm MS}(M))\tau_{\rm cool}(L,M) \;dM
\label{twdlf}
\end{eqnarray}

\noindent where $L$  is the luminosity, $M$ is the  mass of the parent
star (for convenience  all white dwarfs are usually  labelled with the
mass of their main sequence progenitor), $t_{\rm cool}$ is the cooling
time  necessary to  reach a  luminosity $L$,  $\tau_{\rm cool}=dt_{\rm
cool}/dM_{\rm bol}$  is the characteristic cooling  time, $t_{\rm MS}$
is the  main sequence lifetime of  the progenitor of the  white dwarf,
and  $T$ is  the age  of the  population under  study.  The  remaining
quantities,  the  initial  mass  function,  $\Phi(M)$,  and  the  star
formation rate, $\Psi(t)$, are not known  {\sl a priori} and depend on
the physical  properties of the  stellar population under  study.  For
context, excellent fundamental reviews  of how the luminosity function
can  be constructed  for more  general spectral  types and  population
groups can be found in \cite{MB81} and \cite{BM1998}.

Obviously, both the  cooling time and the  characteristic cooling time
must  be  obtained  from  detailed cooling  sequences.   Clearly,  the
cooling rates  are crucial in  determining the white  dwarf luminosity
function.   When  the  characteristic cooling  time  increases  (small
cooling  rates)   the  number   of  white   dwarfs  per   unit  volume
correspondingly  increases.   This  occurs  when either  there  is  an
additional release  of energy in the  core --- such as  the release of
latent heat upon crystallization, see  below --- or when an additional
source  of opacity  in the  atmosphere appears.   The reverse  is also
true.  For instance, when neutrinos are copiously produced in the deep
interior of  the white dwarf  the cooling  rates are large,  hence the
white dwarf  luminosity function  drops below  the nominal  value when
only the contribution due to heat capacity is taken into account.

The  main-sequence lifetime  and a  relation between  the mass  of the
progenitor stars and  the mass of the white dwarf  itself must also be
provided  (this  is known  as  the  initial-final mass  relationship).
Usually, these last quantities are obtained from numerical fits to the
available  pre-white  dwarf evolutionary  sequences  and  also play  a
critical role  in matching observed white  dwarf luminosity functions.
The integration limits  $M_{\rm s}$ and $M_{\rm i}$  play an important
role as  well.  The upper  limit in Eq.~(\ref{twdlf}), is  the maximum
mass for which a  main sequence star is able to  produce a white dwarf
in its final evolutionary stage. As previously mentioned, this mass is
still today  somewhat uncertain.   Theoretical estimates  suggest that
its precise  value is  around $10\,M_{\sun}$  \citep{Rea99}.  Finally,
the lower  limit in Eq.~(\ref{twdlf})  is the  minimum mass of  a main
sequence star  able to produce a  white dwarf of luminosity  $L$ given
the total  age of the  population under study,  and it is  obtained by
solving the following expression:

\begin{equation}
T-t_{\rm cool}(L,M_{\rm i})-t_{\rm MS}(M_{\rm i})=0,
\end{equation}

\noindent which is  to say that the progenitor of  the white dwarf was
born at $t=0$.  Clearly, as  the luminosity decreases the cooling time
increases and  the net result is  that the minimum masses  of the main
sequence  stars able  to  produce  a white  dwarf  of the  appropriate
luminosity  increase.  Thus,  $M_{\rm i}$  approaches $M_{\rm  s}$ for
decreasing luminosities.  This produces a down-turn in the theoretical
white  dwarf  luminosity  function  which   can  be  compared  to  the
observational data, yielding an estimate  of the age of the population
under study.  This has been (and  still is) one of the most successful
applications of the white dwarf  luminosity function.

It is important  to realize that the position of  the down-turn in the
white dwarf luminosity function is  totally independent of the initial
mass function  of the population under  study, a fact that  makes this
method very appealing.  Moreover, the position of the down-turn of the
white dwarf  luminosity function measures  the time elapsed  since the
beginning  of  {\sl  significant   star  formation  activity}  in  the
population under study.  Thus, strictly speaking, only lower limits to
the  age of  the  population can  be retrieved  from  the white  dwarf
luminosity function.  It should also be  noted that the exact shape of
the  down-turn depends  --- although  weakly ---  on the  adopted star
formation rate.  Abrupt down-turns  occur when constant star formation
rates are used, but the slope of the down-turn is shallower for slowly
increasing star formation rates.  Also, beyond the down-turn, there is
a  low-luminosity tail  which is  due to  the contribution  of massive
white dwarfs  and/or contamination  by much  older halo  white dwarfs.
Hence, the shape of this tail is sensitive to the adopted initial mass
function and to the initial-to-final mass relationship. Unfortunately,
present white dwarf  surveys are not nearly deep enough  to reach this
population of ultra-low luminosity white dwarfs.

In  order to  compare  to  the observations  properly,  it is  usually
convenient  to bin  the  theoretical luminosity  function  in one-  or
half-magnitude intervals $\Delta M_{\rm bol}$, in the following way:

\begin{equation}
\langle n(L)\rangle_{\Delta M_{\rm bol}}
=\frac{1}{\Delta M_{\rm bol}}\int_{l-0.5\Delta M_{\rm bol}}^{l +0.5\Delta M_{\rm bol}}
n(L) \;dM_{\rm bol}
\end{equation}

\noindent  where  $\Delta  M_{\rm  bol}$ is  the  size  of  bolometric
magnitude  bin.   This  procedure  introduces  additional  sources  of
uncertainty  that must  be taken  into account  --- see,  for example,
\cite{BBT05}.

\subsection{Monte Carlo simulations}
\label{MC}

The procedure previously described is  the most straightforward way of
constructing   the  theoretical   white  dwarf   luminosity  function.
However,   there  exist   alternatives.    The  key   point  is   that
Eq.~(\ref{twdlf}) does not take into account the many subtle selection
biases that affect the observational  determination of the white dwarf
luminosity  function from  the existing  white dwarf  catalogs.  Monte
Carlo techniques  help to account  for these subtleties.   Even though
prior Monte  Carlo simulators  \citep{WO98, GB99, Tea01,  Tea02, GB04}
adopted very  different approaches,  they demonstrated useful  ways to
evaluate biases in the  observational data.  For instance, \cite{WO98}
distributed white dwarfs according to a previously computed integrated
luminosity function, whereas other Monte Carlo simulators \citep{GB99,
Tea02, GB04, ASPC2013} incorporated  full models of Galactic structure
and  evolution.   Both approaches  are  valid  and produce  reasonable
results  when all  the sample  selection procedures  and observational
biases are taken into account.  Using these tools it has been possible
to  assess the  quality  of the  observational  data, the  statistical
significance of  the samples used  to obtain the observed  white dwarf
luminosity function,  the sample selection procedures,  and the method
used to  derive the white  dwarf luminosity function. We  review these
results below.

Previous    observational   efforts,    which   were    described   in
Sect.~\ref{past}, have  provided an invaluable wealth  of good quality
data.    Moreover,   ongoing   projects   like   those   detailed   in
Sect.~\ref{future}   will   undoubtely    increase   the   sample   of
spectros\-copically-identified    white     dwarfs    with    reliable
determinations of  parallaxes and proper motions,  which are essential
for an accurate determination of  the white dwarf luminosity function.
Last but not least, future space missions like Gaia \citep{PEA01} will
dramatically  increase the  sample  of known  white  dwarfs with  very
accurate astrometric data \citep{Jordan1,Jordan2}.  However, the rapid
increase in  both the quality  and the quantity of  observational data
has  not been  accompanied by  corresponding improvements  in the  way
observational  data are  analyzed.  Thus,  the main  aim of  the Monte
Carlo  simulations  performed  up  to  now  has  been  to  assess  the
reliability of the most common method  used to estimate the disk white
dwarf  luminosity function  --- the  $1/\mathcal{V}_{\rm max}$  method
\citep{S68,S75,F76} --- and to  test other techniques which eventually
could allow more accurate determinations of the white dwarf luminosity
function.  Examples  of these  more sophisticated techniques  are, for
instance, the C$^-$ method  \citep{L71}, the STY method \citep{STY79},
the  Choloniewski   method  \citep{C86},  and  the   Stepwise  Maximum
Likelihood method \citep{E88} which,  among others, are currently used
to derive galaxy luminosity functions.

Two preliminary  studies \citep{WO98,GB99} demonstrated ---  using two
independent Monte  Carlo simulators  --- that  the $1/\mathcal{V}_{\rm
max}$  method for  proper-motion selected  samples is  a good  density
estimator.    However,  it   is  subject   to  important   statistical
fluctuations when estimating the slope of  the bright end of the white
dwarf luminosity function where the space  density of stars is low and
subject to small  number statistical uncertainties.  In  the latter of
these works it was  also shown that a bias in the  derived ages of the
solar  neighborhood  is  present  --- a  consequence  of  the  binning
procedure.  Additionally,  it has  been shown \citep{GTGB06}  that the
size   of    the   observational   error   bars    assigned   by   the
$1/\mathcal{V}_{\rm max}$  method is severely underestimated  and that
more  robust luminosity  function  estimators should  be used.   These
alternative estimators provide a good characterization of the shape of
the white dwarf luminosity function even when small numbers of objects
are used.  Moreover, \cite{GTGB06} found  that for a small sample size
the $1/\mathcal{V}_{\rm max}$ method  provides a poor characterization
of the less  populated bins, while for large  samples the Choloniewski
method and  the $1/\mathcal{V}_{\rm  max}$ method are  comparable.  In
this case  both provide the shape  of the disc white  dwarf luminosity
function and  the precise  location of  the down-turn  with reasonable
accuracy.  This study also showed  that a reliable observational white
dwarf luminosity function can be  best obtained by using a combination
of  both the  $1/\mathcal{V}_{\rm  max}$ method  and the  Choloniewski
method, while  other parametric maximum-likelihood estimators  are not
recommended  for small  sample  sizes.  These  preliminary tests  also
showed  that these  sophisticated  algorithms work  better for  larger
sample  sizes. Undoubtely,  with the  advent of  massive sets  of good
quality observational  data employing  these algorithms will  be among
the  priorities of  the research  field.  More  recently \cite{TGBI07}
have shown that when the sample size  is small it might be affected by
the Lutz-Kelker bias  \citep{LK73} and that contamination  of the disk
sample by high--velocity  halo white dwarfs can  have dramatic effects
on the low--luminosity  bins. Work in this direction will  also be one
of the priorities during the next few years.


\section{Overview of the white dwarf cooling theory}
\label{cooling}

As mentioned  previously, the  key ingredient to  computed theoretical
white  dwarf luminosity  functions is  a detailed  description of  the
cooling  process of  white  dwarfs.   In this  section  we provide  an
overview of  the procedure.   The evolution  of a  carbon-oxygen white
dwarf from the planetary nebula  phase to its disappearance depends on
the properties  of the envelope and  the core.  This process  has been
discussed in  detail in  a large number  of papers  \citep{IT84, KS86,
DM89, W92,  Sea94, BA99,  H99, Cea00, Sea00,  FBB01, PMS02}  --- see
also \cite{IAGB13} for a relatively brief review of the cooling theory
of carbon-oxygen white dwarfs.  The cooling of helium white dwarfs has
also  received extensive  attention \citep{BA98,  HP98, Dea98,  Dea99,
SEG00, Aa01a, Aa01b, Sea01, Sea02}, whereas the cooling of oxygen-neon
white  dwarfs has  received less  attention \citep{One1,  One2, One3}.
Independently of  the chemical  composition of  the core,  the cooling
process can  be roughly  divided into  four stages:  neutrino cooling,
cooling in  the fluid  phase, crystallization  and Debye  cooling.  We
discuss each phase  in the following subsections. In  what follows the
luminosity (instead of  the magnitude) will be used  to describe these
phases, as it is customary in  the field.  To ease the comparison with
observational studies we remind the reader that the absolute magnitude
and the  luminosity of the white  dwarf are related by  the well-known
expression $M_v=-2.5\log(L/L_{\sun})+4.74+ {\rm  B.C.}$, where B.C. is
the bolometric correction that accounts  for the portion of the star's
spectral energy distribution that does not pass through the $V$ filter
\citep{1973asqu.book.....A}.

\subsection{The cooling phases}
\label{phases}

The first evolutionary  phase of typical white dwarfs  is dominated by
neutrino cooling.  This phase occurs for $\log (L/L_{\sun}) \ga -1.2$.
It is  rather complicated because  the initial conditions of  the star
and the behavior  of the envelope are still not  yet fully understood.
For instance, if  the thickness of the hydrogen layer  is large enough
the luminosity due to hydrogen  burning through the pp-chain may never
stop and  could become dominant  at low luminosities, i.e.,  $-3.5 \la
\log (L/L_{\sun})  \la -1.5$  \citep{IT84}. In  this case  the cooling
rate would be similar to that obtained ignoring this source of energy,
and it would be observationally  impossible to distinguish between the
two.  However, the importance  of continued H-burning strongly depends
on the  mass, $M_{\rm H}$, of  the hydrogen layer.  If  $M_{\rm H} \le
10^{-4}\,  M_{\sun}$,  the pp  contribution  quickly  drops and  never
becomes dominant,  except for  low-mass, low-metallicity  white dwarfs
\citep{enuc, Camissasa}.  Since astero-seismological observations seem
to constrain the  size of $M_{\rm H}$ well below  this critical value,
this  source can  be neglected.   Fortunately, when  neutrino emission
becomes  dominant,  the different  thermal  structures  converge to  a
unique  solution,  assuring  the   uniformity  of  models  with  $\log
(L/L_{\sun})\la-1.5$.  Furthermore, since the  time necessary to reach
this value is $\la 8\times 10^7$ years for any model \citep{DM89}, its
influence in the total cooling time is negligible.

After  the neutrino  cooling  phase the  core of  the  white dwarf  is
liquid.  This  stage occurs  for luminosities in  the range  $-1.5 \ga
\log (L/L_{\sun}) \ga -3$.  Here gravothermal energy becomes dominant.
Since  the plasma  is not  very strongly  coupled, its  properties are
reasonably well  known \citep{Sea94}.  To characterize  the properties
of the plasma it is customary to define the Coulomb coupling parameter

\begin{equation}
\Gamma=\langle  Z^{5/3}\rangle\Gamma_{\rm e}
\label{coulomb}
\end{equation}

\noindent where

\begin{equation}
\Gamma_{\rm e}=\frac{e^2}{a_{\rm e}k_{\rm B}T},
\end{equation}

\noindent   $Z$   is  the   atomic   charge,   $a_{\rm  e}$   is   the
inter-electronic distance, $k_{\rm B}$  is the Boltzmann constant, $e$
is the  electron charge, and  $T$ is  the temperature.  This  phase is
characterized  by  small  Coulomb coupling  parameters,  $\Gamma<179$.
During this  phase the  observed luminosity is  controlled by  a thick
non-degenerate  layer  with  an  opacity  dominated  by  hydrogen  (if
present) and helium, and it is  weakly dependent on the metal content.
The main source of uncertainty is related to the chemical structure of
the interior,  which depends on  the rather uncertain adopted  rate of
the \cago  reaction and on  the treatment given to  semiconvection and
overshooting during the pre-white  dwarf evolutionary phases.  If this
rate is high, the oxygen abundance is higher in the center than in the
outer layers.  This results in a reduction of the specific heat at the
central layers where the oxygen abundance  can reach values as high as
$X_{\rm O}=0.85$ \citep{SA97}.

When the Coulomb coupling parameter  reaches a critical value, $\Gamma
\simeq 179$, crystallization at the center of the white dwarf sets in,
and a  new cooling phase  starts, as early recognized  by \cite{KD60},
\cite{AA60}, and \cite{SA61}.  Crystallization leads to the release of
latent  heat,  which  then  controls the  evolution  of  white  dwarfs
\citep{VH68,LH75}, and to  the release of gravitational  energy due to
phase  separation.  The  typical  luminosities during  this phase  are
$\log  (L/L_{\sun})  \la  -3$.  The release  of  gravitational  energy
associated  with  changes  in  the  chemical  composition  induced  by
crystallization  in  carbon-oxygen  mixtures   has  been  examined  by
\cite{ST80,MO83,phaseD}.  Finally, the  consequences of the deposition
of  $^{22}$Ne, the  most abundant  of  the impurities  present in  the
central regions of  a white dwarf have been  examined \citep{IE91}.  A
similar calculation for $^{56}$Fe, the second most important impurity,
was made somewhat later \citep{XV92}.   We elaborate on this below, in
Sect.~\ref{minor}.   Crystallization  introduces  two new  sources  of
energy: latent heat and sedimentation  (a form of gravitational energy
release).  In the  case of Coulomb plasmas, the latent  heat is small,
of the order  of $k_{\rm B} T_{\rm s}$ per  nucleon, where $T_{\rm s}$
is the temperature  of solidification.  Its contribution  to the total
luminosity is small, $\sim 5$\%, but not negligible \citep{SK76}.

During   the  crystallization   process,   the  equilibrium   chemical
compositions  of   the  solid  and  liquid  plasmas   are  not  equal.
Therefore, if  the resulting solid  flakes are denser than  the liquid
mixture, they sink  towards the central regions. If  they are lighter,
they rise upwards and  melt when the solidification temperature, which
depends  on the  density,  becomes  equal to  that  of the  isothermal
core. The  net effect is a  migration of the  heavier elements towards
the  central  regions with  the  subsequent  release of  gravitational
energy  \cite{MO83}.  The  efficiency of  the process  depends  on the
detailed chemical composition and on the initial chemical profile.  It
is most  efficient in a  mixture made of  half oxygen and  half carbon
uniformly distributed throughout the star.

The first calculation  of a phase diagram for C/O  mixtures yielded an
eutectic shape \citep{ST80},  denoting a mixture of  elements in fixed
proportions that solidifies  and melts at a given  temperature that is
lower  than the  melting  points of  either  constituent or  different
mixture.   This  resulted  from  the assumption  that  the  solid  was
entirely random, so that the free energy is given by $F\sim-0.9\Gamma$
\citep{Sea94},  where $\Gamma$  is  defined  by Eq.~(\ref{coulomb})  .
Since the mixture  retains some short range order, the  free energy is
then given  by the  linear mixing rule,  $F_{\rm lm}  \sim -0.9\langle
Z^{5/3}\rangle\Gamma_{\rm e}$.   The solid  phase is less  stable, and
thus $ F>F_{\rm  lm}$, resulting in an eutectic behavior  of the phase
diagram.   The density  functional  theory of  freezing  and the  mean
spherical  approximation,  but with  the  same  diameter for  the  two
chemical species, were  used later to compute  the correlation between
the  particles, and  a  phase  diagram of  spindle  form was  obtained
\citep{BH88}.  A similar  calculation in the framework  of the density
functional   theory,  but   using  the   Improved  Hypernetted   Chain
approximation  to compute  the correlation  functions, was  performed,
producing  an  azeotropic  phase diagram  \citep{II88},  indicating  a
mixture  of liquids  whose proportions  are  not affected  by a  phase
change.   Finally, these  calculations  were extended  to include  the
effects  of  the  different  diameters of  the  two  chemical  species
\citep{SC93} and  the same  results of  \cite{BH88} were  found.  More
recently, \cite{Horowitz}  computed a  new phase  diagram for  the C/O
mixture using an advanced  technique, direct molecular dynamics, again
agreeing with previous results, thus settling this issue.

The  so-called Debye  cooling phase  comes  last. It  occurs at  small
luminosities, typically  $\log(L/L_{\sun})\la -4.5$, when the  star is
almost entirely crystallized.   During this cooling phase  for a large
fraction of  the crystallized core  the specific heat  follows Debye's
law, and scales  as $T^3$.  Due to the reduced  specific heat, cooling
accelerates and  the star's  luminosity drops abruptly.   However, the
outer layers still have very  large temperatures relative to the Debye
temperature,  and  since their  total  heat  capacity is  still  large
enough, they prevent  the sudden disappearance of the  white dwarf, at
least when the envelope is thick \citep{DM89}.

\subsection{The opacity of the envelope}
\label{opacity}

The importance of the atmospheric  treatment in the cooling models for
white dwarfs  cannot be  overstated. In  the strongly  degenerate core
energy transfer by electron conduction dominates.  This has been shown
to be a very efficient mechanism.  Thus, the cores of white dwarfs are
essentially isothermal.   However, in the outer  layers radiation and,
depending  on  the  effective temperature,  convection  dominates  the
energy  transfer.    In  these  layers  the   temperature  profile  is
determined by  the equation  of state.   Thus, changes  in atmospheric
parameters directly affect the core temperature and, since white dwarf
cooling is driven largely by the slow leakage of the thermal reservoir
stored in the core, it moderates  the cooling rate. Until recently the
atmospheric treatment was based on grey atmospheres and Rosseland mean
opacities.    Modern   calculations  incorporate   detailed   non-grey
atmospheres.    Recall    that   at   low    effective   temperatures,
collision-induced absorption  by molecular hydrogen due  to collisions
with H$_2$  represents a major source  of opacity in the  infrared and
dominates the shape  of the emergent spectrum.  Thus,  very cool white
dwarfs with hydrogen-dominated atmospheres begin to turn blue at about
4,500 K,  whereas helium-dominated  atmospheres resemble  black bodies
\citep{H98,SJ99,R01}.   The  cooling  rate,  in  both  cases,  depends
sensitively on the  adopted mass and composition of  the envelope, and
the  age differences  are  substantial, larger  than  1.5~Gyr, at  the
relevant luminosities, $\log(L/L_{\sun})\la-4.5$ \citep{H98}.

We note here that for all the cooling phases described in the previous
section  the   importance  of   the  envelope   is  crucial   for  two
reasons. Clearly,  more transparent envelopes result  in faster energy
losses and shorter cooling times.  This is true for all the previously
listed  cooling phases.   The second  reason  is more  subtle and  was
largely  overlooked  until some  time  ago  \citep{FBB01}, but  it  is
essential  to  take into  account  during  the crystallization  phase.
Theoretical  calculations  predict  that  at  approximately  the  same
evolutionary  stage in  which  crystallization sets  in, the  external
convection  zone penetrates  the  region where  thermal conduction  by
degenerate electrons is very efficient.   Such an occurrence (known as
convective  coupling), initially  produces a  further decrease  in the
cooling rate, followed  by a more rapid decline. In  fact, it has been
proven that the delay introduced by  the convective coupling can be as
large   as   that   produced  by   chemical   differentiation   during
crystallization \citep{FBB01}.

\subsection{The role of minor chemical species}
\label{minor}

Minor chemical  species like $^{22}$Ne  or $^{56}$Fe can also  play an
important role in  the cooling of white dwarfs.   These minor chemical
species are the  products of the pre-white  dwarf evolutionary stages.
The most abundant is $^{22}$Ne.   Its abundance is directly related to
the initial  abundances of  CNO elements,  which, after  the H-burning
phase become $^{14}$N.  This isotope, in turn, becomes $^{22}$Ne after
the  series  of  reactions  $^{14}$N$(\alpha,  \gamma)^{18}$O$(\alpha,
\gamma) ^{22}$Ne, during  the He-burning phase.  Because  of its large
neutron number  and the high  sensitivity of degenerate  structures to
the electron number  profile, $^{22}$Ne can induce a  large release of
gravitational  energy  if, as  a  consequence  of crystallization,  it
migrates towards the center  during crystallization \citep{IE91}.  For
stars  of solar  metallicity, the  typical abundances  are 1--2\%.   A
similar effect can  be produced by the deposition of  $^{56}$Fe at the
center \citep{XV92}.  In this case, typical abundances are 0.1\%.

The physics of the deposition of  the minor species is intricate since
it depends on  the behavior of a multicomponent phase  diagram that is
not yet  known. A first step  consists in assuming that  the C/O/Ne or
C/O/Fe mixtures behave as an effective binary mixture composed of neon
(or iron) and  an average element, representative of  the C/O mixture.
The phase  diagram for arbitrary ionic  mixtures as a function  of the
charge ratio has been computed by  \cite{SC93}.  It was found that the
phase diagram is of the spindle form for $0.72 \la Z_1/Z_2 <1$, of the
azeotropic form  for $0.58 \la Z_1/Z_2  \la 0.72$ and of  the eutectic
form for $  Z_1/Z_2 \la 0.58$.  In  the case of a C/O  mixture made of
equal  mass fractions  of  carbon and  oxygen,  the resulting  average
element has such  a charge that the corresponding  phase diagram shows
an azeotropic  behavior with an  azeotropic abundance of $X_{\rm  a} =
0.16$.  This means that white dwarfs  are in the neon-poor side of the
phase diagram.  Consequently, the solid in equilibrium with the liquid
has a smaller concentration of neon  and, since it is lighter than the
surrounding liquid, it will rise and  melt in lower density regions so
that  the neon  concentration in  the  liquid will  increase until  it
reaches the azeotropic composition.   This process will continue until
all  $^{22}$Ne  is collected  in  a  central  sphere of  mass  $M_{\rm
WD}X_0({\rm  Ne})/X_{\rm  a}  ({\rm Ne})$.   Following  the  procedure
described in  \cite{IEA97} for computing  the decrease in  the cooling
rates suggests that the total energy  release in the case of $^{22}$Ne
is   $\Delta   E\simeq   1.52\times   10^{47}$~erg   for   a   typical
$0.6\,M_{\sun}$ white  dwarf.  At  the corresponding  luminosity, this
would result in  an unrealistic delay of about  9~Gyr, indicating that
there is  a problem  with the  adopted assumptions.   For the  case of
$^{56}$Fe  under  the same  conditions  the  energy released  is  much
smaller,  $\Delta  E\simeq  2.0\times   10^{46}$  erg,  and  thus  the
resulting time delay is also much smaller, $\Delta t\simeq 1.1$~Gyr.

The assumption of an effective binary mixture of the C/O/Ne mixture is
probably not  very realistic. In  fact, a preliminary  ternary diagram
has been  computed for  the C/O/Ne  mixture \citep{SE96}.   This phase
diagram displays  the expected behavior  at the binary  limit (spindle
form for  the C/O mixture,  azeotropic form  for the C/Ne  mixture and
spindle form for the O/Ne mixture).  For small concentrations of neon,
of  the order  of  a  few percent,  and  temperatures  well above  the
azeotropic temperature, the crystallization  diagram is not influenced
by the presence  of neon.  However, as the  temperature approaches the
azeotrope, the resulting solid is  lighter than the surrounding liquid
and the distillation process starts as in the previous case.  The main
difference  is that  it  starts in  the outer  layers  instead of  the
central  layers  and  the  effect  of  separation  is  therefore  much
smaller. As a  matter of fact, the total energy  released in this case
is  $\Delta  E\simeq 0.20\times  10^{46}$  erg  and the  corresponding
effect on  the cooling times are  hence much smaller, of  the order of
0.6 Gyr.

Up to this  moment we have reviewed the role  played by the impurities
in   the   cooling   of   white  dwarfs   during   the   crystallizing
phase. However,  it is  important to realize  that $^{22}$Ne  can also
play an important  role during the liquid cooling  phase.  Building on
earlier   work    \citep{Bea92}   it    has   been    recently   shown
\citep{BH01,DB02,GB08,A10}  that due  to the  large neutron  excess of
$^{22}$Ne  it sinks  towards the  interior as  the liquid  white dwarf
cools.  The subsequent gravitational energy released slows the cooling
of the  white dwarf  by 0.25--1.6  Gyr by the  time it  has completely
crystallized, depending  on the  white dwarf mass  and on  the adopted
sedimentation rate.   This effect  will make  massive white  dwarfs or
those in metal-rich  clusters appear younger than their  true age.  It
has been demonstrated  \citep{GB10} that this is indeed  the case, and
that the slowdown of the white dwarf cooling rate owing to the release
of gravitational energy from $^{22}$Ne sedimentation and carbon-oxygen
phase separation upon crystallization  is of fundamental importance to
reconcile the age discrepancy of the very old, metal-rich open cluster
NGC~6791.  Nevertheless, although the  white dwarf luminosity function
of this  open cluster  provides a statistical  measure, a  direct test
remains to be  done. Since there is no way  to measure the metallicity
of single white dwarf progenitors we would need a wide binary composed
of a non-evolved star and a white dwarf, for which we could accurately
measure   the   age   and  metallicity   using   independent   methods
\citep{Zhao}.

\subsection{Uncertainties in the cooling ages}
\label{uncertainties}

\begin{table*}[]
\caption{Uncertainties in  the estimates of the cooling  time of white
         dwarfs.
         \label{tab:cool}}
\begin{center}
\small
\begin{tabular}{@{}lcl}
\hline
\hline
Input                  &  $\Delta t$ (Gyr) &  Comments              \\ 
\hline
Core composition       &  $\la 0.6$        &  Depending on the \cago\\
Opacity                &  $\la 0.4$        &                        \\
Conductive opacity     &  $\la 1.0$        &                        \\
Metals in the envelope &  $\approx 0.2 $   &                        \\ 
Metallicity of the progenitor & $\la 0.2$  &                        \\
\hline
\multicolumn{3}{c}{Additive contributions of the crystallization 
		   process}                                         \\ 
\hline
C/O                    &  0.8--1.2         & Depending on the \cago \\ 
Fe                     & $\la 1.3$         &                        \\ 
Ne                     & $\la 0.5$         & Ternary mixture        \\
\hline
Observational          &   1--2            &                        \\
\hline
\hline
\end{tabular}
\end{center}
\end{table*}

Table~\ref{tab:cool} displays the uncertainties  in the time necessary
for a typical white dwarf of $0.6\, M_{\sun}$ to reach a luminosity of
$\log(L/L_{\sun}) =  -4.5$.  In the  bottom section of this  table the
additive  contributions  to the  uncertainty  due  to the  physics  of
crystallization  are  shown, whereas  the  top  section describes  the
uncertainties  due  to the  rest  of  the  input physics.   The  major
contribution  is  provided  by  the minor  chemical  species  and  all
contributions are  of the same  order of magnitude, $\sim  1$~Gyr. The
largest  contributions   come  from  the  core   composition  and  the
conductive   opacities  \citep{PMS02,   PMS07}.    Also  the   adopted
composition  of  the  envelope  has  an  important  influence  on  age
estimates for a white dwarf population, although hydrogen-rich cooling
sequences are usually adopted, as most observational determinations of
the white dwarf luminosity function rely on the population of DA white
dwarfs.

There is, however, another point  of concern, namely how the different
numerical  implementations  of  the   evolutionary  codes  affect  the
accuracy  of  the cooling  times.  This  issue  has been  examined  by
\cite{Maurizio}.   These authors  compare  the  cooling ages  obtained
using a set  of controlled input physics and  very different numerical
schemes.  This comparison shows that when the same physical inputs are
adopted the  cooling ages  do not  differ by more  than a  very modest
$\sim 2\%$ at  all luminosities, in sharp contrast  with main sequence
ages, for  which the  typical differences  are of  the order  of $\sim
6\%$,  or   even  larger.   This   difference  is  smaller   than  the
uncertainties   in  cooling   times   attributable   to  the   present
uncertainties in the white dwarf chemical stratification. Hence, white
dwarf cooling ages turn out to  be even more robust than main sequence
evolutionary ages.


\section{Other key ingredients}
\label{inputs}

In this section we discuss the other main quantities needed to compute
theoretical  white  dwarf  luminosity functions.   In  particular,  we
discuss   the  appropriate   choice   of   an  initial-to-final   mass
relationship, which main  sequence lifetimes can be  used to calculate
luminosity  functions,  which  initial   mass  functions  are  usually
employed in the theoretical calculations, and finally, we also provide
a preliminary discussion of the role of the star formation rate.

\subsection{Initial-to-final mass relationship}
\label{IFMR}

One of  the key  ingredients of a  theoretical white  dwarf luminosity
functions is  a relationship linking the  mass of the white  dwarf and
that of its progenitor at the zero age main sequence. It is well known
that   theoretical  evolutionary   calculations  predict   a  positive
correlation between  both quantities.   However, the precise  shape of
the initial-to-final  mass relationship  is troublesome to  predict on
theoretical  grounds alone.  This is  a consequence  of the  intrinsic
difficulties  associated   with  modeling  several   complex  physical
phenomena involved in the final phases of stellar evolution.  A number
of  theoretical   works  have   attempted  to  address   these  issues
\citep{Inma, Girardi2000,  Marigo, Karakas, Marigo2,  Salaris09, R10},
but the area is an ongoing active field of research, and remains open.
In  general, the  theoretical predictions  depend sensitively  on many
subtle  details  of   the  evolutionary  codes  \citep{Weidemann2000}.
However, there is a general consensus that this relationship is almost
linear for progenitor masses between $\simeq 1.2$ and $6.5\, M_{\sun}$
\citep{Kalirai2008, Catalan2008a, Cea08, Casewell2009}, although there
are indications that the slope  becomes steeper for masses larger than
roughly $3.5 \, M_{\sun}$ \citep{Dobbie2009}.  The question of whether
this relationship depends  on the metallicity of  the population under
study remains  controversial ---  see, for  instance \cite{Camissasa},
and references therein.

Much effort has been invested  in empirically determining the slope of
the  initial-to-final  mass relationship.   This  is  usually done  by
employing either  open clusters or detached,  non-interacting binaries
composed of a  white dwarf and a main sequence  star. The best-studied
systems of the latter kind are  common proper motion pairs.  For these
binaries   we   are  confident   that   both   stars  are   physically
associated. In both  cases we know the total age,  and hence the total
age  of a  given white  dwarf (cooling  age plus  the lifetime  of its
progenitor). This allows determination of  the mass of its progenitor,
once  the mass  of  the  white dwarf  is  measured,  provided that  an
accurate relationship between the mass  and the main sequence lifetime
is  available ---  that is,  a set  of reliable  isochrones. The  main
drawback  of   employing  open   clusters  for   this  task   is  that
well-populated clusters are needed.   Additionally, since usually open
clusters are young, they only probe a limited range of masses, between
2.5  and  $7.0\, M_{\sun}$,  an  interval  corresponding to  the  most
massive  white   dwarfs.   Moreover,   open  clusters  often   show  a
metallicity spread, and  the information about the  metallicity of the
progenitor of  the white dwarf is  lost once the star  becomes a white
dwarf.      This,    in     turn,    introduces     an    uncertainty.
\cite{Weidemann1977}   pioneered   the    observational   efforts   to
empirically  determine the  initial-to-final  mass relationship  using
open clusters, studying white dwarfs in Hyades and Pleiades. Following
this    early   work    several    other    clusters   were    studied
\citep{Weidemann1987,    Weidemann2000,   Ferrario2005,    Dobbie2006,
Kalirai2008, Rubin2008,  Casewell2009, Dobbie2009,  Williams2009}, and
it is  expected that this work  will be extended to  fainter and older
clusters in the coming years.

Common  proper  motion  pairs   \citep{Wegner1973,  OHL88}  provide  a
work-around for  most of these  concerns.  In such binaries  the stars
have never interacted and both  components evolve as single stars.  As
they were born simultaneously, the total age of the white dwarf can be
split into its  cooling age and that of its  progenitor star.  The age
of the companion  star (hence, the total age of  the binary system) is
measured using an independent method, for example a set of theoretical
isochrones of the appropriate metallicity, from rotation rate, or from
chromospheric activity.  The main inconvenience of this method is that
reliable age determinations are available for only a handful of binary
systems \citep{Catalan2008a, Cea08, Zhao}.

A  third method,  though less  frequently employed,  to determine  the
shape of  the initial-to-final mass  relationship is to  calculate the
difference  of cooling  times between  two  white dwarfs  in a  double
degenerate  system \citep{Finley}.   Until  recently  this method  was
hampered   by    the   small    number   of   such    systems   known.
\cite{2015ApJ...815...63A} identified new candidate double degenerates
in the  SDSS, bringing the total  known to 142.  For  over 50 systems,
they were  able to  derive masses  and cooling  ages from  Balmer line
spectra and  employed a Bayesian  statistical approach to  fitting the
most probable  initial-to-final mass ratio consistent  with the sample
for initial masses of $2-4\,  M_{\sun}$.  Open clusters provide little
leverage  within this  mass  range, so  double  degenerates provide  a
valuable alternative approach.   Since these methods make  use of both
observed  data  and  theoretical  models  the  results  are  known  as
semi-empirical initial-to-final mass relationships.

In any case, there are two  options to compute theoretical white dwarf
luminosity   functions.    Either   semi-empirical,   or   theoretical
initial-to-final mass relationships can be employed.  In general, most
theoretical  calculations of  the  luminosity function  of disk  white
dwarfs  are  done  using semi-empirical  relationships,  whereas  when
luminosity  functions of  stellar systems  with known  metallicity are
computed  theoretical  initial-to-final   mass  relationships  of  the
appropriate metallicity  are employed.  This includes  the calculation
of white dwarf luminosity functions for the Galactic halo, or for open
and globular clusters in the Galaxy.

\subsection{Main sequence lifetimes}
\label{tms}

Main  sequence lifetimes  are also  necessary to  theoretically derive
white dwarf  luminosity functions --- see  Eq.~(\ref{twdlf}). Ideally,
one should  use main  sequence lifetimes  consistent with  the adopted
initial-to-final  mass  relationship.   However, this  is  not  always
possible.   Hence,  many  calculations employ  a  simple  relationship
between the mass  of the progenitor of  a white dwarf at  the zero age
main sequence and its lifetime. An example of this simple relationship
is that of \cite{IL89}:

\begin{equation}
t_{\rm MS}=10^{10} \left(\frac{M}{M_{\sun}}\right)^{-3.5}\,{\rm yr}
\end{equation}

However, there  are more  sophisticated treatments, which  include the
use of  interpolation in theoretical  isochrones.  An example  of such
more elaborated  treatments consists in interpolating  within the {\tt
BaSTI}    isochrones\footnote{\url{http://albione.oa-teramo.inaf.it/}}
for  the  appropriate  metallicity   of  the  white  dwarf  progenitor
\citep{Pietrinferni}.

\subsection{The initial mass function}
\label{IMF}

Another,  less relevant  but necessary,  input to  compute theoretical
white dwarf  luminosity functions is  the initial mass  function. This
function peaks at a few tenths of  a solar mass, and shows an extended
tail for masses larger than this value.  It is customary to model this
tail as  a power law,  with a  fixed exponent, or  in some cases  as a
combination  of power  laws, again  with fixed  exponents. A  thorough
discussion of  the initial mass  function is  far beyond the  scope of
this work,  but the interested  reader will find useful  the excellent
review of \cite{Kroupa2002}. As a  matter of fact, most studies employ
the classical Salpeter-like  initial mass function \citep{Salpeter55},
which  is  a  power  law  with just  one  index  $\alpha$.   That  is,
$\Phi(M)\propto M^{-\alpha}$,  where $M$ is  the mass at the  zero age
main sequence.  In some cases the power-law index is considered a free
parameter  to  fit  the  observations.   However,  more  sophisticated
studies prefer to employ the  so-called ``universal'' mass function of
\cite{Kroupa2001}  ---  which for  the  mass  range relevant  to  most
stellar systems is  totally equivalent to a two-branch  power law with
exponent $-\alpha$, with $\alpha = 1.3$ for $0.08 \leq M/M_{\sun}<0.5$
and $\alpha  = 2.30$ for  $M/M_{\sun} \geq 0.5$. Finally,  sometimes a
top-heavy initial mass function:
\begin{equation}
\Phi(M) = \frac{1}{M} \exp\left(\frac{-\log(M/\mu)}{2\sigma^2}\right)
\end{equation}
is adopted.  In this expression $\mu=10\, M_{\sun}$ and $\sigma=0.44$.
This initial mass  function was introduced by  \cite{Suda2013}, and is
dominated by  high mass stars.  It  has been found that  this function
better  reproduces  the  characteristics  of  metal-poor  populations,
namely those with [Fe/H]$\leq-2$.

\subsection{The star formation rate}
\label{SFR}

The final ingredient  in the calculation of a  theoretical white dwarf
luminosity function  is the star  formation rate.  The  star formation
rate depends on the characteristics of the population under study. For
the case of the disk white  dwarf population a constant star formation
rate is usually adopted, whereas for metal-poor populations --- namely
the stellar spheroid and the system  of globular clusters --- a short,
intense burst of star formation is adopted. Typically, the duration of
the burst  is a  free parameter  that can  be used  to better  fit the
observed   white    dwarf   luminosity   function   and    the   color
distribution. In fact, if all the ingredients to model the white dwarf
luminosity function are known with  good accuracy one could eventually
use Eq.~\ref{twdlf} to  solve the inverse problem and  derive the star
formation history  of the population under  study. Unfortunately, this
is not  usually possible,  and moreover, the  solution of  the inverse
problem does  not have  a unique  solution.  We  elaborate on  this in
Sect.~\ref{SFH}.


\section{Applications of the white dwarf luminosity function}
\label{applications}

\subsection{The white dwarf population and the age of the 
            Galaxy}
\label{properties}

As  mentioned  above,  the  potential  use of  white  dwarf  stars  as
chronometers  was recognized  several decades  ago \citep{Sch59},  but
only in  the last two decades  has there been good  observational data
and  reliable theoretical  cooling  sequences which  are necessary  to
interpret  them  in  terms  of  an  age  for  the  Galactic  disk.   A
determination from the  white dwarf luminosity function of  the age of
the halo population  still remains to be done, due  to the scarcity of
halo white dwarfs in the solar neighborhood.

The evidence  for a decrease in  the number counts of  white dwarfs at
lower  luminosities  was  suspected   but  unobtainable  until  larger
telescopes  with  more quantum  efficient  detectors  were applied  to
systematic surveys for  white dwarfs.  In fact, the  early white dwarf
luminosity functions were constrained by  the available data, and only
the bright portion  of the luminosity function ---  namely, those bins
with       $\log(L/L_{\sun})\la-3.5$       ---      had       reliable
determinations. However,  the improved observational  efforts resulted
in  the first  convincing identification  of a  down-turn in  the disk
white dwarf  luminosity function \citep{LDM88}, which  was interpreted
as  a  consequence of  the  finite  Galactic age  \citep{WA87}.   This
discovery was  followed by a  series of papers in  which progressively
more  sophisticated   cooling  sequences,  observing   strategies  and
analysis techniques were employed \citep{GB88, W92, HA94, OSW96, GB96,
BRL97, R00}.

The effect  of the white  dwarf initial-to-final mass relation  on the
luminosity  function  has  been  examined  by  \cite{Cea08}.   Fitting
theoretical   luminosity  functions   from   a   variety  of   stellar
evolutionary  inputs  to  the   average  of  the  observed  luminosity
functions   of    \cite{LDM88},   \cite{OSW96},    \cite{LRB98},   and
\cite{KHH99},  it  was  shown  that the  only  significant  difference
between  the  theoretical fits  occurs  beyond  the down-turn  in  the
luminosity function,  where current data  is sparse.  They  also found
that  a disk  age of  $\sim 11$~Gyr  consistently fit  their composite
observed luminosity function best, a value that is substantially older
than prior determinations.   At present there does not appear  to be a
generally-accepted age  for the Galactic  disk derived from  the white
dwarf luminosity function --- it should still be regarded as a work in
progress  because  of  the  difficulty in  obtaining  a  statistically
significant sample of stars in  the most important faintest luminosity
bins  as  well as  remaining  uncertainties  in core  composition  and
atmospheric opacities in  the coolest white dwarfs.   However, a rough
average derived from  the references discussed in this  review --- see
Table~\ref{spaced}  ---  suggests that  star  formation  in the  solar
neighborhood (thin  disk) began  about 10~Gyr  ago.  This  estimate is
uncertain by at least ten percent.

Finally, it is worth emphasizing that the age estimates obtained using
the down-turn in the disk  white dwarf luminosity function, regardless
of the  observed sample, are  very robust lower limits.   As mentioned
earlier, the white cooling ages obtained employing different numerical
evolutionary codes differ by only a few percent when a standard set of
physical assumptions is adopted, a quite remarkable feature.  Not only
that, it  has been recently  demonstrated \citep{Fe-H14} that  the age
estimates of the  Solar neighborhood obtained by  fitting the position
of the  down-turn in the disk  white dwarf luminosity function  do not
depend  on  the  adopted  metallicity   law,  also  ensuring  the  age
determined in this way very reliable.  However, a study of super-solar
metallicity  stars,  [M/H]~ $>$~+0.1   dex,  by  \cite{Kordopatis2015}
employing data from the RAdial Velocity Experiment (RAVE) DR4 suggests
that the  angular momentum  of numerous stars  have been  increased by
scattering at  corotation resonance of  the Galaxy's spiral  arms from
regions well  within the Sun's  Galactocentric radius.  This  may have
ramifications  for  the local  white  dwarf  luminosity function  that
remain to be explored.

One would  expect the  age of  the halo derived  from the  white dwarf
luminosity function  to be more  uncertain than that obtained  for the
disk, primarily because  the halo component is older  and comprises at
most a  few percent of  the local  white dwarf population.   Thus, the
typical halo white dwarf is distant  and faint.  However, based on the
handful of currently available studies of the halo white dwarfs in the
solar   neighborhood   and   in    nearby   globular   clusters   (see
Table~\ref{spaced}) it can  be said that star formation  began in that
population  subgroup at  least  11~Gyr ago.   

\subsection{The star formation history of the Galaxy}
\label{SFH}

In addition to an absolute age for the Galactic disk and space density
of white  dwarfs, the  white dwarf  luminosity function  also contains
information about the star formation  and death rates over the history
of the  Galaxy.  However, it  is important  to note that  because cool
white  dwarfs  have  very  long evolutionary  time  scales,  the  past
Galactic  star  formation  rate  influences   the  shape  of  the  low
luminosity portion of the white  dwarf luminosity function.  Also, due
to the extremely long main sequence lifetimes of low mass stars, which
are  the progenitors  of bright  white dwarfs,  the shape  of the  hot
branch of the white dwarf luminosity function is also sensitive to the
past star  formation activity.   All this implies  that the  past star
formation  activity  is  still  influencing the  present  white  dwarf
birthrate and  that the  past star formation  rate could  be retrieved
from  the  white  dwarf  luminosity   function,  as  clearly  seen  in
Eq.~(\ref{twdlf}).  Provided that we  have a reliable determination of
the white dwarf  luminosity function and accurate  cooling models, the
star formation rate, $\Psi(t)$, can  be obtained by solving an inverse
problem.   This  has  not  been  possible  yet  due  in  part  to  the
discrepancies  between the  different observational  determinations of
the white dwarf  luminosity function and also to  the still relatively
large error bars in each bin.  In addition, from the theoretical point
of view, the inverse transformation  cannot be easily done because the
kernel of the transformation is  complicated and, more importantly, it
is not  symmetric.  These  are the  reasons why  very few  papers have
explored the  possibility of  doing some  sort of  Galactic archaelogy
using the luminosity function of white  dwarfs as a tracer of the star
formation activity.  However, this situation may be improving --- see,
for example, \cite{RH11}.

In order  to overcome  these difficulties, several  possibilities have
been  suggested. The  first and  most straightforward  method requires
{\sl a  priori} knowledge of the  shape of the star  formation history
and  consists of  adopting a  trial function  that depends  on several
parameters, followed  by a search  for the values of  these parameters
that best fit the observed  luminosity function.  This is accomplished
by  minimizing  the  differences  between the  observed  and  computed
luminosity functions \citep{Y89,NS90,I95,I01}.  The second possibility
consists of computing the luminosity function for massive white dwarfs
\citep{DEA94},  which have  negligible main  sequence lifetimes,  thus
making much easier the solution of the inverse problem. Unfortunately,
such white dwarfs are rare. Also, if a significant fraction of massive
white  dwarfs results  from double  white dwarf  mergers, the  problem
becomes more complicated, as the  method relies on the assumption that
white dwarfs with moderately high masses  (say between 0.8 and $1.1 \,
M_{\sun}$)  are  the  result   of  single  star  evolution.   Finally,
\cite{R13}  presented recently  an algorithm  for inverting  the white
dwarf luminosity function  to obtain a maximum  likelihood estimate of
the star  formation rate in  the solar neighborhood. As  expected from
the discussion above,  the results were found to be  most sensitive to
the choice of white dwarf cooling  models. Use of the algorithm on two
independent determinations of the white dwarf luminosity function gave
similar  results:  a bimodal  star  formation  rate with  broad  peaks
2--3~Gyr  and  7--9~Gyr  before   the  present,  with  star  formation
commencing   about   8--10~Gyr  ago.    \cite{Tremblay2014}   employed
individual white dwarf atmosphere models and a complete volume-limited
``20 parsec sample''  to investigate the local  star formation history
in the  Solar neighborhood and  concluded that an enhancement  in star
formation rate (and consequent increase  in white dwarf space density)
occurred within the  past 5~Gyr, enhancing the space  density of white
dwarfs by  a factor of  $\sim 2.5$.  This  roughly agrees with  a peak
reported by \cite{R13}.

Several  attempts have  been  made to  discern  structure in  observed
luminosity functions.  An early  study \citep{NS90} modeled the effect
that various  bursts of star formation  would have on the  white dwarf
luminosity function and suggested a  modest bump noted by \cite{LDM88}
might be evidence for such an event  about 0.3~Gyr ago.  The impact of
merger episodes in the Galactic disk on the white dwarf population was
examined by  \cite{Tea01} using a  Monte Carlo simulator.   This study
concluded that only  relatively small merger episodes  involving a few
percent  or less  of  the  current disk  population  white dwarfs  are
compatible with the current kinematics  of known white dwarfs and that
the white  dwarf luminosity  function is  insensitive to  such events.
More recently,  attention has been  called to  a plateau in  the white
dwarf luminosity function  obtained from the Sloan  Digital Sky Survey
near $M_{\rm  bol}=10.5$ \citep{Hea06} ---  see Figure~\ref{fig:disk}.
This luminosity corresponds to a cooling  time of 0.3~ Gyr.  Adding the
main-sequence lifetime of 2.5 Gyr~for the typical progenitor suggests
that a drop in  star formation rate may have occurred  about 3 Gyr ago
after a burst  or long duration of higher rate  of star formation.  So
far, not much other work appears  to have been attempted in this area,
probably because until  very recently the sample size  of white dwarfs
have  been small  enough that  statistical uncertainties  dominate the
luminosity bins.  However, there is reason to be hopeful, because much
larger samples of white dwarfs are becoming available.

Of  particular interest  are  the  delays between  the  onset of  star
formation  in the  thin disk,  thick disk  and halo.   However, a  few
caveats are in  order.  Some of these include possible  changes in the
initial mass function  \citep{AL96,GM97,BCR99}, uncertainties in white
dwarf core composition  and chemical profile \citep{IEA97,SA97,PAB00},
phase  separation  \citep{IEA97,Mo99},  incompleteness of  the  sample
\citep{MR01,HOS02},      unresolved      binary     star      fraction
\citep{LBH05,FBZ05} and  statistical limitations of the  method chosen
to construct the white dwarf luminosity function \citep{GB99,GTGB06}.

A number of  studies have been published which  attempted to determine
the properties of the halo \citep{MA90,TA90,IE98b}. It is important to
realize that the halo white dwarf luminosity function not only carries
information  about the  age of  the halo  but also  --- under  certain
circumstances --- about the initial mass  function of the halo. If, as
generally accepted, the halo was formed in a very short time scale the
halo  star formation  rate  can be  well approximated  by  a burst  of
negligible   duration  and   the  calculation   of  the   integral  of
Eq.~(\ref{twdlf}) can be  simplified, since for all  halo white dwarfs
we have

\begin{equation}
T=t_{\rm MS}(M)+t_{\rm cool}(l,M)
\end{equation}

In this scenario,  each luminosity corresponds to a given  mass of the
white  dwarf  progenitor.   In  other  words,  the  halo  white  dwarf
luminosity function maps the mass of the progenitor of the white dwarf
as a  function of the  luminosity, $M=M(l)$. Taking into  account that
the white dwarf luminosity function is  the number of white dwarfs per
unit bolometric magnitude, $n(l)\propto dN/dl$, we have:

\begin{equation}
n(l)\propto\frac{dN}{dl}=\frac{dN}{dM}\frac{dM}{dl}=\Phi(M)\frac{dM}{dl}
\end{equation}

Thus, once we have complete samples of the halo white dwarf population
and  reliable observational  determinations  of the  halo white  dwarf
luminosity  function, the  initial mass  function of  the halo  can be
experimentally  obtained.   This  assumes,  of course,  that  we  have
accurate and precise white dwarf cooling sequences. Another concern is
that the initial-to-final mass relation  for white dwarfs could depend
upon  the  metallicity of  their  main  sequence progenitors  ---  see
\cite{Zhao}. The expectation is that low metallicity halo white dwarfs
are born  with thicker hydrogen  envelopes, leading to  more prolonged
shell burning than in disc white dwarfs.  In addition, stars of higher
primordial metallicity are  expected to be more  efficient in shedding
mass  during the  AGB phase,  resulting  in lower  mass white  dwarfs.
Consequently, this  could affect  the white dwarf  luminosity function
and   its  interpretation   for  different   populations  \citep{enuc,
Camissasa}.   Fortunately,  the  most  recent  work  on  this  problem
suggests that  the shape of  the white dwarf luminosity  function from
bright  to   faint  bin  is  relatively   insensitive  to  metallicity
\citep{Fe-H14}.

\subsection{Other applications}
\label{other}

Other  applications of  the  white dwarf  luminosity function  include
independent constraints  on the  physical mechanisms  operating during
the  cooling process,  an independent  test  of the  constancy of  the
gravitational constant, $G$, and its  use as an astro-particle physics
laboratory. We examine all three below.

As  shown  previously, the  white  dwarf  luminosity function  carries
important information  about the physics of  cooling.  Consequently, a
reliable white dwarf  luminosity function would allow us  to test both
the mechanisms operating at  high effective temperatures --- basically
neutrino  cooling  ---  and  the mechanisms  which  are  dominant  for
relatively low core temperatures  (crystallization). Neutrinos are the
dominant  form of  energy  loss in  model white  dwarf  stars down  to
$\log(L/L_{\sun})  \simeq  -2.0$,  depending   on  the  stellar  mass.
Consequently,  the evolutionary  timescales of  white dwarfs  at these
luminosities sensitively  depend on the  ratio of the  neutrino energy
loss to the photon  energy loss. Thus, the slope of  the bright end of
the white  dwarf luminosity function directly  reflects the importance
of  neutrino emission.   Although  the unified  electroweak theory  of
lepton  interactions  that  is   crucial  for  understanding  neutrino
production has been well tested in the high-energy regime --- see, for
instance,  \cite{H97} for  an  excellent review  ---  the white  dwarf
luminosity function  could help to  test the low-energy regime  of the
theory   \citep{Wea04,   TGB05}.    For  example,   recent   work   by
\cite{dipole}  using   the  best  available  white   dwarf  luminosity
functions  and  including  the  effects of  observational  errors  and
binning have set a firm limit  on the neutrino magnetic dipole moment;
$\mu_\nu<5\times 10^{-12}e  \hbar/(2m_{\rm e}c)$.  This  is comparable
to the constraints on $\mu_\nu$ set by studies of globular clusters.

The white  dwarf luminosity function  can also put constraints  on the
physical  mechanisms that  operate  at low  core temperatures,  namely
crystallization and  phase separation.   As discussed  previously, the
inclusion  of phase  separation  upon crystallization  in the  cooling
sequences  adds  an  additional  delay  to  the  cooling  (and,  thus,
considerably  modifies   the  characteristic  cooling  times   at  low
luminosities),  which depends  on  the initial  chemical profile  (the
ratio of carbon  to oxygen) and on the transparency  of the insulating
envelope.  Thus, if  a direct measure of the disk  age with reasonable
precision is obtained by an  independent method, say via main sequence
turn-off stars,  the white  dwarf luminosity function  directly probes
the physics  of crystallization. It is  worth noting as well  that not
only  the exact  location of  the down-turn  of the  disk white  dwarf
luminosity  function  is  affected  by  the  details  of  the  cooling
sequences but  also the position and  the shape of the  maximum of the
white dwarf luminosity function.  Thus, additional tests are possible.

A second application  of the white dwarf  luminosity function involves
setting constraints  on a hypothetical variation  of the gravitational
constant, $G$.   There are  two reasons for  this.  First,  when white
dwarfs are cool enough, their  energy is entirely of gravitational and
thermal origin, and any change in the value of $G$ modifies the energy
balance.  This in turn translates into a change of luminosity. Second,
since they are long-lived objects, $\sim 10$ Gyr, even extremely small
values of the  rate of change of $G$ can  have detectable effects. The
first attempts to  obtain constraints on $\dot G$ from  the cooling of
white  dwarfs \citep{Vila76}  were  unsuccessful due  to  the lack  of
reliable  observational  data and  the  uncertainties  in the  cooling
theory of  white dwarfs.  Since  then both the observational  data and
the cooling theory  have been improved substantially,  as discussed in
Sect.  \ref{cooling}. It has been shown \citep{GB95} that for the case
of a  secularly varying  $G$, the  luminosity of  a cool  enough white
dwarf is given by:

\begin{equation}
  L=-\frac{dB}{dt}+\frac{\dot G}{G}\Omega
  \label{LG}
\end{equation}

\noindent where $B=U+\Omega$ is the binding energy, $U$ is the thermal
energy and  $\Omega$ is the  gravitational energy. As the  white dwarf
cools,  the  thermal   content  decreases  and  the   second  term  in
Eq.~(\ref{LG}) dominates.  Note as  well that  the cooling  process is
accelerated  if  $\dot G/G  <0$.   By  comparing  the results  of  the
previous equation with  the observed position of the  down-turn in the
white  dwarf  luminosity  function,   the  following  firm  limit  was
obtained:

\begin{equation}
-(1\pm 1) \times 10^{-11}\; \mbox{yr}^{-1} < \frac{\dot G}{G} <0 
\end{equation}

\noindent at the $1\sigma$  confidence level \citep{GB95}. This result
was challenged  by \cite{torres},  who obtained  a much  tighter bound
using the same method, but their analysis was subsequently shown to be
flawed.  This  issue  was  finally settled  by  detailed  evolutionary
calculations  \citep{gnew,JCAP},  which corroborated  the  preliminary
analytical calculations of \cite{GB95}. We mention here that a tighter
bound was obtained  by \cite{GB6791} using the  white dwarf luminosity
function of the open, metal-rich, well-populated cluster NGC~6791:

\begin{equation}
-1.8 \times 10^{-12}\; \mbox{yr}^{-1} < \frac{\dot G}{G} <0 
\end{equation}

The last  and most  exotic application of  the white  dwarf luminosity
function we would like to discuss here is its use as an astro-particle
physics laboratory.  Pulsating white  dwarfs have been used frequently
for such  a purpose \citep{FBB01}.   For example, the  plasma neutrino
process  has been  tested  using  pulsating DB  white  dwarfs ---  see
\cite{Wea04}, and  references therein. The existence  of neutrinos has
been known for  many years, yet there are other  exotic particles that
have been postulated by theorists that  have not been detected so far.
The  white dwarf  luminosity function  can help  in this  regard.  For
instance, the  mass of  the axion has  been constrained  using ZZ~Ceti
white  dwarfs \citep{Cea01,Iea10,M3B+I}.   The white  dwarf luminosity
function  appears to  leave  little room  for  other theorized  weakly
interacting  exotic particles  such  as ``massive  dark photons´´  and
``dark sector particles'' lighter than a few keV \citep{DI,UI}. It has
also been shown very recently that the white dwarf luminosity function
can  be used  to derive  consistent upper  limits to  the mass  of the
axion,  and the  axion-electron  coupling constant  ($g_{\rm ae}$)  of
DFSZ-axions \citep{Iea08}.   Clearly, more accurate  constraints would
be enabled by improved observational white dwarf luminosity functions.

\section{The future}
\label{future}

Although  it has  been  used  to construct  nearly  every white  dwarf
luminosity function  to date, the $1/\mathcal{V}_{\rm  max}$ technique
is  very  vulnerable to  undetected  incompleteness  and small  sample
statistical uncertainties.  It  has been shown ---  see, for instance,
Sect.~\ref{MC}, \cite{WO98}, and \cite{GB99} --- that the age estimate
resulting  from  the $1/\mathcal{V}_{\rm  max}$  method  is also  very
sensitive to the choice of assumed Galactic disk scale height, binning
interval,  placement of  bin centers,  and that  the space  density is
inherently uncertain by at least  50\% in samples containing less than
$\sim 100$  stars.  Also, the vast  majority of known white  dwarfs do
not meet the  completeness magnitude and proper  motion limits imposed
by the $1/\mathcal{V}_{\rm max}$ technique, so they cannot be included
in the  standard white dwarf  luminosity function.  It has  been shown
\citep{GTGB06}  that other  techniques  not previously  used on  white
dwarfs do better in identifying the luminosity down-turn and the total
space  density.   It  will  be important  to  use  more  sophisticated
techniques such  as these on  the large  samples of white  dwarfs that
will be available soon.

The next decade  holds much promise for a  definitive determination of
the  white dwarf  luminosity  function  for each  of  the three  major
stellar  components  of  the  solar  neighborhood.   Among  the  first
opportunities for  improvement will  be PanSTARRS  data \citep{Kea02},
which will cover the entire sky observed  in the course of the SDSS on
timescales of  less than a  week, reaching 24$^{\rm th}$  magnitude in
each  frame ---  2--3  magnitudes  fainter than  the  best large  area
surveys available today.  PanSTARRS opens a new time domain on the sky
that  has  exciting  potential  for  discovery  in  a  wide  range  of
astronomy, ranging from  the search for near earth  asteroids to gamma
ray burster  afterglows.  Two of  the science objectives  of PanSTARRS
will  contribute  directly to  improving  the  white dwarf  luminosity
function.  The Solar Neighborhood (SOL)  project will build an all-sky
parallax  catalog for  stars within  $\sim 100$  pc over  its ten-year
lifetime.  It  is expected to  provide a volume-limited sample  out to
$\sim 50$ pc suitable for studies  of brown dwarf stars and cool white
dwarfs.  The Extragalactic and Galactic Stellar Science (EGGS) project
will  provide a  proper motion  catalog for  $\sim 10^8$  stars, whose
precision reaches $\sim 1 \, {\rm mas/yr}$.  This astrometric database
will be a  goldmine for white dwarf searches, especially  in the halo,
via reduced proper motion  diagrams.  The astrometric grid established
by PanSTARRS also  will comprise a faint object  reference catalog for
higher precision but sparser surveys such as Gaia.

In  a sense,  PanSTARRS  is  a pilot  project  for the  Large-aperture
Synoptic  Survey Telescope  (LSST) project.   The LSST  is a  proposed
ground-based  8.4-meter, 10  square-degree-field  telescope that  will
image  the  entire sky  every  three  nights in  continuous  30-second
exposures.  It will  open a movie-like window on the  sky.  One of the
proposed  by-products  of  the  data stream  is  a  parallax  catalog.
Assuming astrometric  precision of a  few $\mu$as per  observation, it
has been  shown \citep{Ivezic08,Saha09} that short-arc  parallaxes can
be measured  for stars out  to 10 pc in  a few months  of observation.
LSST   multi-band  photometry   will  permit   the  determination   of
photometric parallaxes, chemical abundances and ages via colors at the
turn-off for main-sequence  stars at all distances  within the Galaxy.
With a geometric parallax accuracy  of 1 milli-arcsecond and exposures
reaching $g =  25$, the LSST parallax survey will  match the faint-end
precision  of  planned space-based  missions  like  Gaia, providing  a
complete catalog to at least $M_{\rm v}  = 15$ through the half of the
Galaxy visible from its site in Chile.

One of the  most-anticipated astrometric surveys will  be conducted by
the ESA  satellite Gaia \citep{PEA01}.   It is expected to  yield high
precision parallaxes as well as  proper motions.  The typical accuracy
of  the parallaxes  will  be of  the order  of  26~$\mu$as at  $V=15$,
degrading to 600~$\mu$as at $V=20$.   The measured proper motions will
be good to 0.2~mas per year.  Gaia will provide multi-color photometry
for 1.3 billion  stars at $V =  20$. Also, it was  expected to measure
radial  velocities to  a  few km/s  precision  to at  least  $V =  17$
\citep{Jordan2}, but recently discovered scattered light problems will
probably  reduce the  performance of  radial velocity  spectrometer to
$V=15$.  In any case,  Gaia's spectroscopic instrument is specifically
designed to  work around the near-infrared  Ca triplet and so  will be
useless for  most white dwarfs.   Nevertheless, Gaia will  generate an
unprecedented sample of stars from  which to construct the white dwarf
luminosity  function   for  all   three  components  of   the  Galaxy.
Specifically, it  is foreseen that  Gaia will discover  around 400,000
new white  dwarfs \citep{Jordan1}.  Using a population  synthesis code
\cite{TGB05} showed that the disk white dwarf population can be probed
out  to  at  least  400~pc  and apparent  Gaia  magnitude  $G  =  21$.
Distinguishing between disk and halo  white dwarfs will require fairly
sophisticated  automatic  classification algorithms  \citep{Tea98}  in
addition to the usual reduced proper  motion diagrams.  The age of the
disk and space  density of disk white dwarfs  will be well-determined,
and  hypothetical merger  episodes in  the disk  can be  investigated.
However,  due to  the  poor  sensitivity of  Gaia's  detectors at  red
wavelengths, most likely only the bright  half of the halo white dwarf
luminosity function will be probed \citep{TGB05}.


\section{Summary of age and space density determinations}
\label{space}

\begin{sidewaystable}[]
\caption{Summary  of the  space densities of disk ---  top sections, 
         in units of $10^{-3}$~pc$^{-3}$ --- and halo white dwarfs ---
         bottom section, in  units of $10^{-6}$~pc$^{-3}$ --- obtained
         by different authors.
         \label{spaced}}
\begin{center}
\small
\begin{tabular}{@{}ccccc}
\hline
\hline
Age (Gyr) &
Sample size & 
Source &  
$n$ &
References \\ 
\hline
\multicolumn{5}{c}{Disk (All white dwarfs)}\\
\hline
---                 & 23           & 10 pc sample              & $>5$                   & \cite{SL77}        \\
---                 & 20           & Proper motion             & 7.5                    & \cite{E83}         \\
$9.3\pm2.0$         & 43           & Field white dwarfs        & 3.0                    & \cite{WA87,LDM88}  \\
$9.5^{+1.1}_{-0.8}$ & 50           & Wide binaries             & $5.3^{+3.5}_{-0.7}$    & \cite{OSW96}       \\
$8.0\pm 1.5$        & 43           & \cite{LDM88}              & 3.39                   & \cite{LRB98}       \\
$10.0^{+3}_{-1}$    & 53           & Proper motion             & 4.16                   & \cite{KHH99}       \\
---                 & 43           & \cite{LDM88,LRB98}        & 2.5                    & \cite{MR01}        \\
7.5                 & 46           & 13 pc sample              & $5.0\pm0.7$            & \cite{HOS02,Cea03} \\
---                 & $\sim 6,000$ & SDSS                      & $4.6\pm0.5$            & \cite{Hea06}       \\
---                 & 44           & 13 pc sample              & $4.8\pm0.5$            & \cite{HSO08}       \\ 
---                 & 3,358        & SDSS                      & ---                    & \cite{DG08}        \\
---                 & 8            & Deep field                & $3.46^{+1.71}_{-1.20}$ & \cite{Sea09}       \\
---                 & $\sim 10^4$  & Proper motion             & $3.19\pm0.09$          & \cite{RH11}        \\
\hline
\multicolumn{5}{c}{Disk (Hot white dwarf samples)}\\
\hline
---                 & 353          & Hot PG white dwarfs only  & $0.49\pm0.05$          & \cite{FLG86}       \\
---                 & 41           & Hot AAT white dwarfs only & $0.60\pm0.09$          & \cite{B89}         \\
---                 & 298          & Hot PG white dwarfs only  & $0.50\pm0.05$          & \cite{LBH05}       \\
\hline
\multicolumn{5}{c}{Halo}\\
\hline
---                 & 6            & Proper motion             & $13\pm6$               & \cite{LDM89}       \\
---                 & 2            & Proper motion             & $\sim 700$             & \cite{Iea00}       \\
---                 & 38           & Proper motion             & $\sim 220$             & \cite{Oea01}       \\
---                 & 33           & SDSS                      & $0.4$                  & \cite{Hea06}       \\
11.47               & $\sim 1000$  & Globular clusters         & ---                    & \cite{Hea07}       \\
---                 & $\sim 10^4$  & Proper motion             & $4.4\pm1.3$            & \cite{RH11}        \\
10--11              & 3            & SDSS                      & ---                    & \cite{Kea10}       \\
\hline
\hline
\end{tabular}
\end{center}
\end{sidewaystable}

Here we  summarize the  most relevant  observational estimates  of the
ages   and   space   densities    of   white   dwarfs   discussed   in
Sects.~\ref{past}  and   \ref{future}.   We  do  not   pretend  to  be
exhaustive, but provide an overview  of the results obtained from what
we  consider  to  be  the   most  relevant  observational  data  sets.
Table~\ref{spaced} shows the local white dwarf space densities for the
Galactic disk (top sections) and  halo (bottom section). Also shown in
this table  are the  derived age, when  available (first  column), the
number  of stars  in the  sample  (second column),  the source  (third
column) and the appropriate reference  (last column).  As can be seen,
the  estimated  space  densities  of  disk  white  dwarfs  are  fairly
consistent --- except for the  very initial studies.  A simple average
yields $4.3\times  10^{-3}$~pc$^{-3}$.  This estimate is  uncertain by
at least  10\%.  The estimates of  the white dwarf number  density for
the  halo  population  are  rather discrepant,  but  the  most  recent
estimates indicate that the white dwarf  number density in the halo is
two to three orders of magnitude lower than the disk.

Finally,  we would  like to  mention that  the white  dwarf luminosity
function also  has provided useful independent  age determinations for
several globular  and open  clusters \citep{HGJ95,C95,R98,vH98,vHG00}.
These and other investigations of  the white dwarf luminosity function
in clusters  will not be  discussed in  detail here.  Let  us mention,
however, that clusters provide useful benchmarks for the relative ages
of the  disk and halo, but  they do not  span the entire age  range of
their parent populations.  Each of these components is also subject to
different influences  that can bias the  derived luminosity functions.
In  the  halo,  globular  clusters are  subject  to  uncertainties  in
distance   determinations,  limiting   magnitudes,  abundances,   mass
segregation, tidal stripping, binary interactions, and/or deficiencies
in evolutionary  models \citep{vH98,R00,HS03,HR04}.  In the  disk, the
white dwarf  luminosity functions  of open clusters  sometimes provide
ages  \citep{Beda05} that  sharply  contrast with  those derived  from
main-sequence  turn-off  ages,  if  all the  energy  sources  are  not
properly taken into account \citep{GB10, 6791}.  Clearly there is much
left to be done in this area.


\section{Conclusions}
\label{conclusions}

The field has advanced in several distinct ways since the first robust
white  dwarf   luminosity  function  was   constructed  \citep{LDM88}.
Substantial samples of white dwarfs from  the thin and thick disk have
been assembled.   Convincing samples  of halo  white dwarfs  have been
identified, though they still are too small to allow construction of a
definitive luminosity function.  The  white dwarf luminosity functions
for  open clusters  and  globular clusters  are  beginning to  provide
useful time markers for the early history of the Galaxy.

Probably the most significant improvements during the last decade have
been in the theoretical models for white dwarfs, which now incorporate
more  realistic core  cooling physics  and atmospheric  opacities.  In
addition,  artificial samples  of white  dwarfs have  been constructed
that have helped  quantify the uncertainties in age  and space density
derived  from  the  white  dwarf  luminosity  function,  and  how  the
precision of these important quantities are related to various binning
strategies, sample  sizes, population groups,  selection biases, among
other effects.

Certainly, the next decade will  see a dramatic increase in the number
of known  white dwarfs.  More  than anything else, this  will preserve
the status of  the white dwarf luminosity function as  one of the most
important tools for unraveling the origin and evolution of the Galaxy.


\section*{Acknowledgments}
EG--B acknowledges  partial support  for this  work from  MINECO grant
AYA2014-59084-P, and by the AGAUR.   TDO acknowledges support for this
project from  NSF grants AST-0807919, AST-108845,  and AST-1358787, as
well  as  NASA grant  NNC04GD87G.   The  authors thank  the  anonymous
reviewer,  who   provided  numerous  suggestions   that  substantially
improved this paper.


\bibliographystyle{elsarticle-harv}

\end{document}